\def\singlecol{0}
\def\paper{1}
\if\paper\singlecol
\documentclass[11pt,journal,draftcls,onecolumn]{IEEEtran}
\else
\documentclass[10pt,journal]{IEEEtran}
\fi
\usepackage{amssymb,amsmath,graphicx,psfrag,bm,color,cite,setspace}
\ifCLASSOPTIONcompsoc
\else
\fi

\ifCLASSINFOpdf
\else
\fi

\usepackage{cite}
\usepackage{array}
\usepackage[caption=false,font=footnotesize]{subfig}
\usepackage{fixltx2e}
\usepackage{algorithmic}
\usepackage{todonotes}
\usepackage{amsfonts}
\usepackage{amssymb}
\usepackage{amsthm}
\usepackage{MnSymbol}%
\usepackage{wasysym}%
\usepackage{fancybox}
\usepackage{booktabs}
\DeclareMathOperator*{\tp}{\mbox{\tiny \sf{T}}}

\usepackage{setspace}

\begin{document}

\title{Bounded Influence Propagation $\tau$-Estimation: {\\A New Robust Method for ARMA Model Estimation}}
\author{Michael~Muma,~\IEEEmembership{Member,~IEEE,}
        and~Abdelhak~M.~Zoubir,~\IEEEmembership{Fellow,~IEEE}
\thanks{M. Muma and A.M. Zoubir are with the Signal Processing Group, Technische Universit\"at Darmstadt, Darmstadt, Germany. This work was supported by the project HANDiCAMS which acknowledges the financial support of the Future and Emerging Technologies (FET) programme within the Seventh Framework Programme for Research of the European Commission, under FET-Open grant number: 323944.}}

\maketitle

\if\paper\singlecol
\vspace{-30 pt}
\else

\fi

\begin{abstract}
A new robust and statistically efficient estimator for ARMA models called the bounded influence propagation (BIP) $\tau$-estimator is proposed. The estimator incorporates an auxiliary model, which prevents the propagation of outliers. Strong consistency and asymptotic normality of the estimator for ARMA models that are driven by independently and identically distributed (iid) innovations with symmetric distributions are established. To analyze the infinitesimal effect of outliers on the estimator, the influence function is derived and computed explicitly for an AR(1) model with additive outliers. To obtain estimates for the AR($p$) model, a robust Durbin-Levinson type and a forward-backward algorithm are proposed. An iterative algorithm to robustly obtain ARMA($p$,$q$) parameter estimates is also presented. The problem of finding a robust initialization is addressed, which for orders $p+q>2$ is a non-trivial matter. Numerical experiments are conducted to compare the finite sample performance of the proposed 
estimator to existing robust methodologies for different types of outliers both in terms of average and of worst-case performance, as measured by the maximum bias curve. To illustrate the practical applicability of the proposed estimator, a real-data example of outlier cleaning for R-R interval plots derived from electrocardiographic (ECG) data is considered. The proposed estimator is not limited to biomedical applications, but is also useful in any real-world problem whose observations can be modeled as an ARMA process disturbed by outliers or impulsive noise.
\end{abstract}


\vspace{-5 pt}
\section{Introduction}
\label{sec:intro}

Autoregressive moving-average (ARMA) models are amongst the most popular models for characterizing dependent data and they have a long tradition in numerous real-world applications, e.g. in speech processing \cite{vaseghi-2008}, biomedicine \cite{tarvainen-2004, cassar-2010}, radar \cite{haykin-2007}, electricity consumption forecasting \cite{chakhchoukh-2009c,chakhchoukh-2010t,chakhchoukh-2010j}, system identification \cite{wang2016filtering} and econometry \cite{tsay-2010}. Numerous extensions of the ARMA model, such as Seasonal Integrated ARMA (SARIMA) \cite{chakhchoukh-2010j}, Periodic ARMA (PARMA) \cite{bondon-parma-2015}, Controlled ARMA \cite{ding2014hierarchical}, and Time-Varying ARMA (TV-ARMA) models \cite{ding2006nonstatARMA} have been proposed.

This paper focusses on robust parameter estimation for ARMA models associated with random processes for which the majority of samples are appropriately modeled by a stationary and invertible ARMA model and a minority consists of outliers with respect to the ARMA model. For such cases and, in general, classical estimators are unreliable and may break down completely  \cite{tsay-1988, deutsch-1990, ljung-1993, chen-1993, shin-1996,luna-2001,maronna-2006,muler-2009,zoubir-2012,chakhchoukh-2009c,chakhchoukh-2010t,chakhchoukh-2010j,andrews-2008,louni-2008}. The nature of the outliers depends on the application. For example, motion artifacts are often evident in biomedical signals such as intracranial pressure (ICP), electrocardiographic (ECG) and photoplethysmographic (PPG) signals \cite{han-2013, muma.ssp-2014, strasser-2012, spangl-2007, Schaeck2015} while in electricity consumption forecasting outliers are associated with holidays, major sporting events and strikes \cite{chakhchoukh-2009c,zoubir-2012}. For a 
discussion on how outliers affect ARMA parameter estimation, the reader is referred, e.g. to \cite{deutsch-1990,chen-1993,maronna-2006,Molinares2009,Kharin2011} and there is a clear need for robust methods that can, to some extent, resist outliers. First contributions to robust estimation for dependent data were made in the 1980's \cite{martin-1982,kassam-1985,martin-1986,bustos-1986}, and in recent years, research in this area has increased significantly (e.g. \cite{arce2001,luna-2001,yang-2001,mili-2002,McQuarrie2003,maronna-2006,Chareka2006,aysal-2007,spangl-2007,andrews-2008,louni-2008,liang-2009,muler-2009,chakhchoukh-2009c,nunkesser2009,dong-2010,li-2010,gandhi-2010,chakhchoukh-2010t,chakhchoukh-2010j,strasser-2012,han-2013,becker2014,dehling2015}). 

Research on robust ARMA parameter estimation may be loosely grouped into two categories which are associated with the diagnostic approach (e.g. \cite{tsay-1988, deutsch-1990, ljung-1993, chen-1993, shin-1996,McQuarrie2003,Chareka2006,louni-2008,dehling2015}) and the statistically robust approach (e.g. \cite{luna-2001,maronna-2006,muler-2009,zoubir-2012,chakhchoukh-2009c,chakhchoukh-2010t,chakhchoukh-2010j,spangl-2007,andrews-2008,durre2015,molinari2015}). Diagnostic approaches enhance robustness via detection and hard rejection of outliers, followed by a classical parameter estimation method that handles missing values. Statistically robust methods utilize the entire data set and accommodate the outliers by bounding their influence on the parameter estimates. Robust statistical theory also provides measures, such as the influence function (IF), the breakdown point and the maximum bias curve \cite{maronna-2006, huber-2009, zoubir-2012}, which characterize quantitative and qualitative robustness and allow for 
an analytical comparison of different estimators.

The main contributions of this paper is to propose and analyze a new estimator for ARMA model parameters called the bounded influence propagation (BIP) $\tau$-estimator which is simultaneously robust and possesses a controllable statistical efficiency. Robustness and high efficiency are jointly achieved by incorporating an auxiliary model which prevents the propagation of outliers into the $\tau$-estimator. The term 'propagation of outliers' means that one outlier in the observations creates multiple outliers in the reconstructed innovation series. The BIP $\tau$-estimate minimizes a robust and efficient scale of the reconstructed innovation series. In Theorem 1, strong consistency of the $\tau$-estimator of the ARMA parameters is established. In Lemma 1, Fisher consistency of the $\tau$-estimator of the ARMA parameters is shown, given all past observations. In Lemma 2, almost sure convergence of the $\tau$-estimator of the innovations scale to the population value based on the expectation operator is proven.
 In Theorem 2, under an ARMA model, it is established that the BIP $\tau$-estimator is asymptotically equivalent to a $\tau$-estimator. Theorems 1 and 2 together prove the strong consistency of 
the proposed estimator under general conditions, which include the Gaussian ARMA model as a special case. In Theorem 3, asymptotic normality of the estimator for the ARMA model is proven by deriving the asymptotic equivalence to an M-estimator. To analyze the infinitesimal robustness of the BIP $\tau$-estimator in the asymptotic case, its IF is derived. The IF is explicitly computed for an autoregressive process of order one, AR(1), in the case of additive outliers. To compute the estimates for the AR($p$) model, a computationally efficient robust Durbin-Levinson type algorithm is proposed that incorporates the BIP model. Here the parameters are recursively found for increasing orders. In this way, searching for a robust starting point to minimize a non-convex cost function is avoided, which is a key-difficulty in robust estimation. A forward-backward algorithm to recursively compute the AR($p$) parameters is also proposed. In the search for ARMA parameter estimates, a Marquard algorithm is used to find the 
parameters that 
minimize the $\tau$-scale of the innovations. For this case, an algorithm to find a robust starting point is presented. The starting point algorithm uses a BIP-AR model based outlier cleaning operation. Numerical experiments to evaluate the estimator in terms of the maximum bias curve in order to assess its quantitative robustness and also to compare it to existing benchmark estimators are conducted. In particular, Monte Carlo experiments for ARMA models of orders $4\geq (p+q) \geq 8$ are performed. This is unusual in robust ARMA parameter estimation, which usually is limited to ARMA models of lower orders. Patchy and independent replacement and additive outliers of different types are considered in the simulations. Finally, the proposed estimator is applied to a real-data example of artifact cleaning for R-R interval plots derived from electrocardiographic (ECG) data. R-R intervals denote the time intervals between consecutive heart beats and are used in heart rate and heart rate variability analysis.


{\it Relation to existing work:} In the analysis of our estimator, we build upon theoretical results that were established for the BIP MM-estimator \cite{muler-2009}. As for the classical regression setting, the $\tau$ \cite{yohai-1988} and MM \cite{yohai-1987} are alternative estimators with similar statistical and robustness properties. In the context of AR parameter estimation, a key advantage of the $\tau$-estimator is its definition via the $\tau$-scale. Based on this definition, a robust Durbin-Levinson type procedure is proposed. Further, the starting point for the BIP MM, especially for $p+q>2$ is difficult to find and expressions for the IF are not available for the BIP MM-estimator. Our estimator is also conceptually related to the filtered $\tau$-estimator \cite{maronna-2006}, which uses a robust filter to prevent outlier propagation. A disadvantage of the filtered estimators is that they are intractable in terms of robustness and asymptotic statistical analysis.

{\it The paper is organized as follows.} Section~\ref{sec:signal-model} introduces the signal and outlier models and discusses the propagation of outliers. Section~\ref{sec:prop_est} introduces the BIP $\tau$-estimator and details associated statistical and robustness analysis. Section~\ref{sec:algorithm} presents an algorithm for computing the stationary and invertible BIP $\tau$-estimates. Section~\ref{sec:simulations} compares the performance of the proposed BIP $\tau$-estimator with existing ARMA parameter estimators via Monte Carlo simulations. Section~\ref{sec:real-example} provides a real-data example of artifact cleaning for R-R interval plots derived from ECG data. Conclusions, and possible extensions of this research are presented in Section~\ref{sec:conclusion}.

{\it Notation.} Vectors (matrices) are denoted by bold-faced lowercase (uppercase letters), e.g.  $\mathbf{a}$ ($\mathbf{A}$). The $j$th column vector of a matrix $\mathbf{A}$ is denoted by $\mathbf{a}_j$. $(\cdot)^\top$ is the transpose operator. Sets are denoted by calligraphic letters, e.g. $\mathcal{B}$. $\hat{\boldsymbol{\beta}}$ refers to the estimator (or estimate) of the parameter vector $\boldsymbol{\beta}$, $\log^+(x)=\mathrm{max}(\log(|x|),1)$, $f(x)$ and $F(x)$ are, respectively, the probability density function (pdf) and cumulative distribution function (cdf) of $x$, $f(x_1,x_2)$ and $F(x_1,x_2)$ are, respectively, the joint pdf and joint cdf of the random variables $x_1$ and $x_2$, $ f(x_1|x_0;\phi_1)$ is the pdf of $x_1$ conditioned on $x_0$ and given $\phi=\phi_1$.  $P(x=c)$ is the probability that $x=c$. $\mathrm{E}\left[\cdot\right]$ is the expectation operator, while $\xrightarrow[D]{}\mathcal{N}(\mathbf{0},\boldsymbol{\Sigma})$ denotes convergence to the normal distribution with mean 
vector $\mathbf{0}$ and 
covariance matrix $\boldsymbol{\Sigma}$. Given a function $g(\mathbf{x}):\mathbb{R}^k\rightarrow \mathbb{R}$, $\nabla g(\mathbf{x})$ is the $k$-dimensional column vector whose $i$th element is $\partial g(\boldsymbol{x})/\partial x_i$. Finally, $x_{\mathrm{min}}:\Delta_x:x_{\mathrm{max}}$ denotes the grid of equidistant points in $\mathbb{R}$, ranging from $x_{\mathrm{min}}$ to $x_{\mathrm{max}}$ with a step size of $\Delta_x$. 

\section{Signal and Outlier Models}
\label{sec:signal-model}
The ARMA and Bounded Innovation Propagation (BIP)-ARMA signal models, as well as some important outlier models, are briefly revisited. Attention is drawn to the fact that estimators, which are computed based on the innovations, require a mechanism that prevents the propagation of outliers.

\subsection{Signal model}
Let 
\begin{equation}
\label{eq:obs_upto_t}
\boldsymbol{y}_t = (\ldots, y_{t-k},\ldots,y_{t-1},y_t)
 \end{equation}
denote a sequence of observations that was generated by a stationary and invertible ARMA($p,q$) process up to time $t$ according to
\begin{equation}
\label{eq:y_t_ARMA}
 y_t = \mu_0 +\sum_{i=1}^p \phi_{0,i}(y_{t-i}-\mu_0) + a_t(\boldsymbol{\beta}_0) - \sum_{i=1}^q \theta_{0,i}a_{t-i}(\boldsymbol{\beta}_0)
\end{equation}
where the true parameter vector $\boldsymbol{\beta}_0=(\boldsymbol{\phi}_0,\boldsymbol{\theta}_0,\mu_0)$, $\boldsymbol{\phi}_0=(\phi_{0,1},\ldots,\phi_{0,p})$ and $\boldsymbol{\theta}_0=(\theta_{0,1},\ldots,\theta_{0,q})$.\\
{\bf (A1)} {\it Assume that $a_t$ are independent and identically distributed (iid) random variables with a symmetric distribution and further assume that $\mathrm{E}\left[\log^+(|a_t|)\right]<\infty$.}\\
To restrict the parameter space in a manner which is consistent with a stationary and invertible ARMA model, let $\boldsymbol{\beta}=(\boldsymbol{\phi},\boldsymbol{\theta},\mu)$ be a parameter vector defined by the polynomials 
\begin{equation}
\label{eq:ar_poly_operator}
 \phi(B)=1-\sum_{i=1}^p\phi_iB^i
\end{equation}
and 
\begin{equation}
\label{eq:ma_poly_operator}
 \theta(B)=1-\sum_{i=1}^q\theta_iB^i
\end{equation}
which have all their roots outside the unit circle. Then, by defining
\begin{equation}
\label{eq:arma-inno-poly}
a_t^e(\boldsymbol{\beta})=  \theta^{-1}(B) \phi(B) (y_t-\mu),
\end{equation}
the following recursion follows
\begin{equation}
\label{eq:arma-inno-recursion}
a_t^e(\boldsymbol{\beta})=  y_t - \mu - \sum_{i=1}^p \phi_i(y_{t-i}-\mu)+\sum_{i=1}^q \theta_i a_{t-i}^e(\boldsymbol{\beta})
\end{equation}
and $a_t^e(\boldsymbol{\beta}_0)=a_t$.\\
{\bf (A2)} {\it Assume that $\phi_0(B)$ and $\theta_0(B)$ do not have common roots.}

\subsection{Outlier models}
\label{subsec:outlier_models}
In real-world applications, the observations $y_t$ may not exactly follow (\ref{eq:y_t_ARMA}). There exist several statistical models for outliers in dependent data (see e.g. \cite{tsay-1988, deutsch-1990, ljung-1993, chen-1993, shin-1996,louni-2008,maronna-2006, zoubir-2012}). The following provides a brief review of important models.\\
The {\it additive outlier (AO) model} defines contaminated observations $y^\varepsilon_t$ according to
 \begin{equation}
 \label{eq:contaminated_arma_ao}
y^\varepsilon_t=x_t+ \xi_t^\varepsilon w_t,
 \end{equation}
where $x_t$ follows an ARMA model, as given in (\ref{eq:y_t_ARMA}), $w_t$ defines the contaminating process that is independent of $x_t$ and $\xi_t^\varepsilon$ is a stationary random process for which
 \begin{equation}
 \label{eq:xi}
   \xi_t^\varepsilon= \begin{cases} 1 & \quad \mathrm{\text{with probability }} \varepsilon \\ 0 & \quad  \mathrm{\text{with probability }} (1-\varepsilon). \end{cases}
 \end{equation}
For the {\it replacement outlier (RO) model}
 \begin{equation}
 \label{eq:contaminated_arma_ro}
y^\varepsilon_t=(1-\xi_t^\varepsilon)x_t+ \xi_t^\varepsilon w_t,
 \end{equation}
where $w_t$ is independent of $x_t$ and $\xi_t$ is defined by (\ref{eq:xi}). 
As discussed, e.g. in \cite{maronna-2006,zoubir-2012}, {\it innovation outliers}, i.e., outliers in $a_t$, can be dealt with by classical robust estimators.

Outliers may also differ in their temporal structure. For {\it isolated outliers}, $\xi_t^\varepsilon$ takes the value 1, such that at least one non-outlying observation is between two outliers (e.g. $\xi_t^\varepsilon$ follows an independent Bernoulli distribution).  For {\it patchy outliers}, on the other hand, $\xi_t^\varepsilon, \{ t\in 1,\ldots,n\}$ takes the value 1 for $n_{\mathrm{patch}}\leq n / 2$ subsequent samples. 

\subsection{Bounded innovation propagation (BIP)-ARMA model} 
\label{subsec:bip-arma-model}
ARMA parameter estimation, i.e., determining $\hat{\boldsymbol{\beta}}$, is often based on minimizing some function of the reconstructed innovation sequence. However, as can be seen from \eqref{eq:arma-inno-poly}, one AO or RO in $y_t$ can propagate onto multiple innovations $a_t^e(\boldsymbol{\beta})$. In the extreme case, all entries of the innovations sequence are disturbed by a single outlier. Thus, robust estimators are only applicable if they are combined with a mechanism to prevent outlier propagation. An auxiliary model to do this, is the BIP-ARMA model \cite{muler-2009}:
\if\paper\singlecol
\begin{eqnarray}
\label{eq:bip-arma-model}
y_t = a_t +\mu + \sum_{i=1}^p\phi_i(y_{t-i}-\mu)-\sum_{i=1}^r\bigg(  \phi_i a_{t-i}+(\theta_i-\phi_i)\sigma\eta\left(\frac{a_{t-i}}{\sigma}\right)\bigg)
\end{eqnarray}
\else
\begin{eqnarray}
\label{eq:bip-arma-model}
y_t\! =\! a_t \! +\mu \! + \sum_{i=1}^p\! \phi_i(y_{t-i}-\mu)\! -\! \sum_{i=1}^r\! \bigg(  \phi_i a_{t-i}+(\theta_i-\phi_i)\sigma\eta\left(\frac{a_{t-i}}{\sigma}\right)\! \bigg)\nonumber\!\!\!\!\!\!\!\!\!\!\!\!\!\!\!\!\! \\
\end{eqnarray}
\fi
Here, $r=\mathrm{\text{max}}(p,q)$, where if $r>p$, $a_{p+1}=\ldots=a_r=0$, while if $r>q$, $b_{q+1}=\ldots=b_r=0$. ARMA models are included by setting $\eta(x)=x$. Thus, by choosing $\eta(x)$ to be one of the well-known monotone or redescending nonlinearities (e.g., Huber's or Tukey's) \cite{huber-2009}, all innovations that lie within some region around $\mu$ are left untouched and, on the other hand, the effect of a single AO or RO is bounded to a single corrupted innovation. In \eqref{eq:bip-arma-model}, $\sigma$ is a robust M-scale of $a_t$ \cite{huber-2009, zoubir-2012}, i.e., it solves 
\begin{equation}
 \label{eq:m_scale-functional}
\mathrm{E}\left[\rho\left(\frac{a_t}{\sigma} \right) \right]=b,
\end{equation}
where $b$ is defined as
\begin{equation}
\label{eq:b_Mscale}
 b=\mathrm{E}\left[\rho(x)\right].
\end{equation}
To make the M-estimator consistent in scale with the standard deviation when the data is Gaussian, $\mathrm{E}\left[\cdot\right]$ in \eqref{eq:b_Mscale}, is the expectation operator with respect to the standard normal distribution. \\
{\bf (A3)} {\it Assume that $\rho(x)$ is a real-valued function with the following properties: $\rho(0)=0, \rho(x)=\rho(-x)$, and $\rho(x)$ is continuous, non-constant and non-decreasing in $|x|$. $\psi(x)=\frac{d\rho(x)}{dx}$ is bounded and continuous.} \\
{\bf (A4)} {\it Assume that $\eta(x)$ is an odd, bounded and continuous function.}\\ 
From (\ref{eq:bip-arma-model}), the innovations sequence can be recursively obtained for $t\geq p+1$ according to
\if\paper\singlecol
\begin{equation}
\label{eq:bip-arma-model-inno-recursion}
a_t^b(\boldsymbol{\beta},\sigma)=y_t-\mu-\sum_{i=1}^p\phi_i(y_{t-i}-\mu)+\sum_{i=1}^r\bigg(\phi_i a_{t-i}^b(\boldsymbol{\beta},\sigma)+(\theta_i-\phi_i)\sigma \eta\left(\frac{a_{t-i}^b(\boldsymbol{\beta},\sigma)}{\sigma} \right) \bigg).
\end{equation}
\else
\begin{eqnarray}
\label{eq:bip-arma-model-inno-recursion}
a_t^b(\boldsymbol{\beta},\sigma)&=&y_t-\mu-\sum_{i=1}^p\phi_i(y_{t-i}-\mu)+\sum_{i=1}^r\bigg(\phi_i a_{t-i}^b(\boldsymbol{\beta},\sigma) \nonumber \\ 
& & +(\theta_i-\phi_i)\sigma \eta\left(\frac{a_{t-i}^b(\boldsymbol{\beta},\sigma)}{\sigma} \right) \bigg).
\end{eqnarray}
\fi
Fig.~\ref{fig:dependent-arma21-inno-est} illustrates the influence of $\eta(\cdot)$ for an ARMA(2,1) model with parameters $\boldsymbol{\phi}_0 = (-0.39,-0.3)$, $\theta_0= 0.9$. The red crosses mark the AO positions in the observations. When reconstructing the innovations with an ARMA model (\ref{eq:arma-inno-recursion}) that uses $\boldsymbol{\beta}_0$, multiple innovation samples are contaminated. This effect is suppressed when applying the BIP-ARMA (\ref{eq:bip-arma-model-inno-recursion}).
\if\paper\singlecol
 \begin{figure}[htp]
    \centering
   \includegraphics[width=0.55\textwidth]{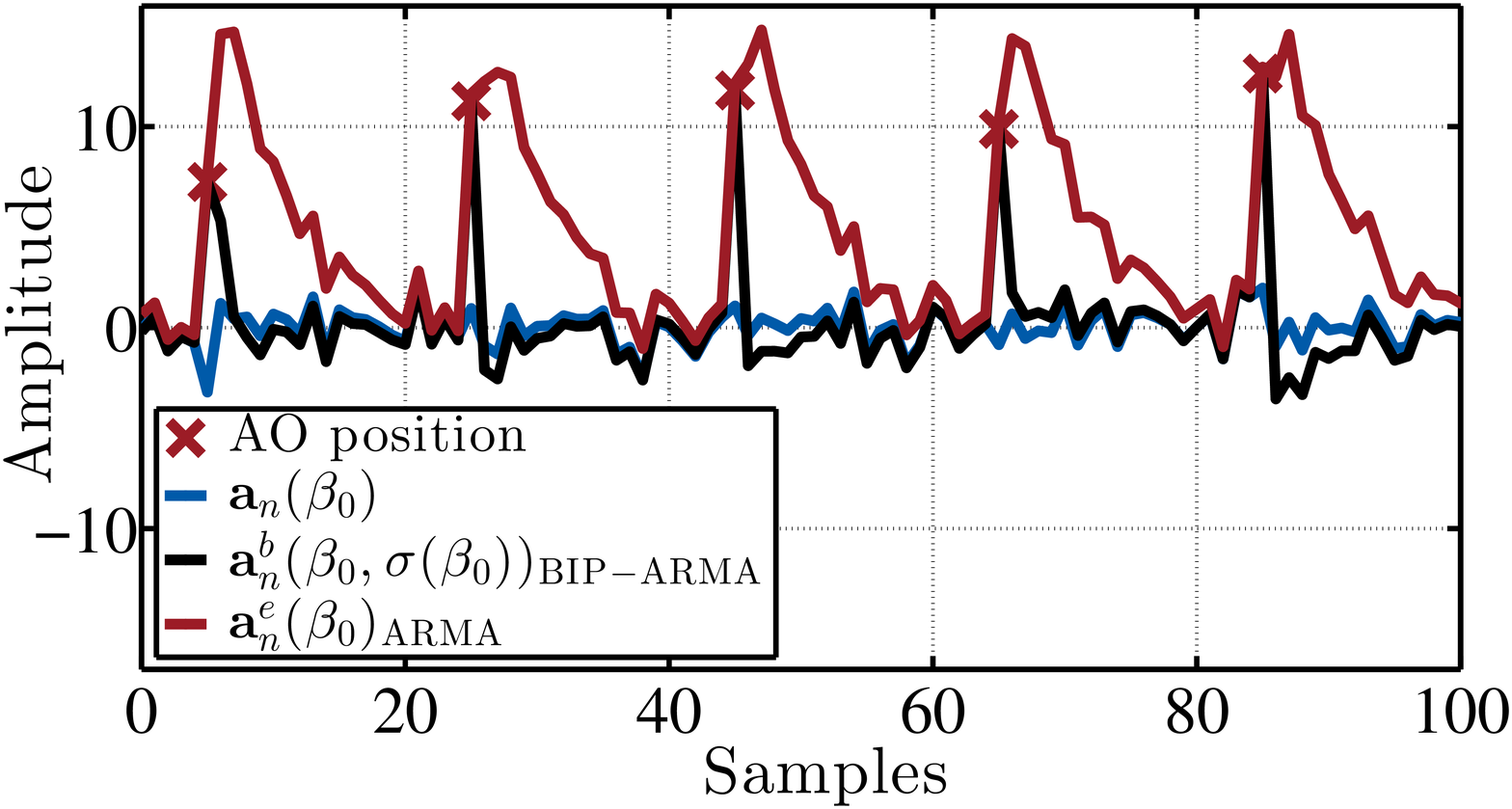}  
      \caption{(blue) True innovations sequence; (red) innovations derived from a Gaussian ARMA(2,1) observation with AOs whose positions are marked with red crosses; (black) innovations obtained when using a BIP-ARMA(2,1) model. In both cases, the true parameter vector $\boldsymbol{\beta}_0$, is used.}
      \label{fig:dependent-arma21-inno-est}
\end{figure}
\else
 \begin{figure}[htp]
    \centering
   \includegraphics[width=0.5\textwidth]{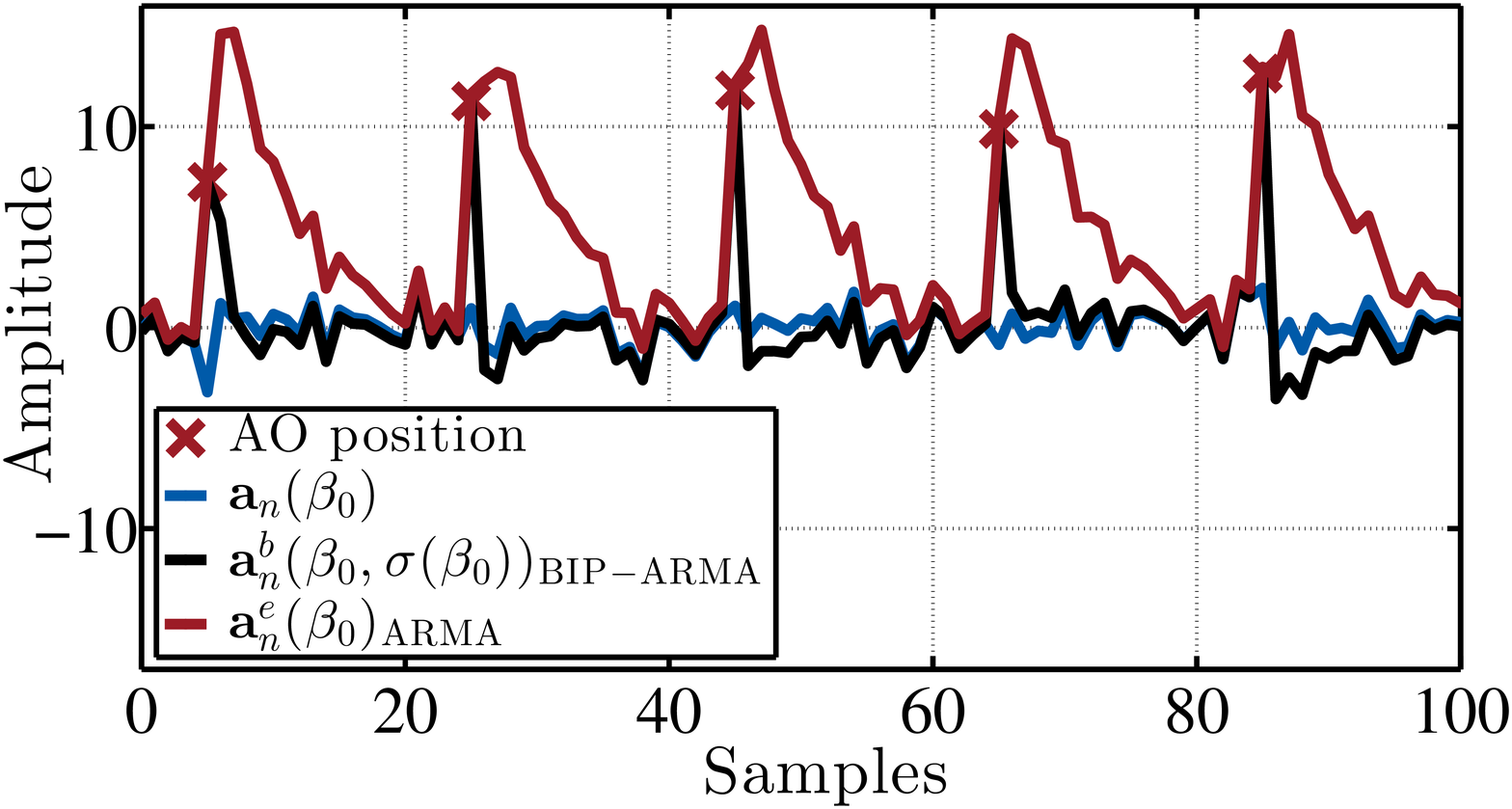}  
  \vspace{- 20 pt}
      \caption{(blue) True innovations sequence; (red) innovations derived from a Gaussian ARMA(2,1) observation with AOs whose positions are marked with red crosses; (black) innovations obtained when using a BIP-ARMA(2,1) model. In both cases, the true parameter vector $\boldsymbol{\beta}_0$, is used.}
      \label{fig:dependent-arma21-inno-est}
\end{figure}

\fi
\vspace{-5 pt}
\section{Proposed Estimator}
\label{sec:prop_est}
We next define an estimator that is based on the idea of minimizing a robust and efficient scale of the reconstructed innovations, the $\tau$-scale. The estimator is defined for the case when the sample size exceeds the number of model parameters, i.e., $n>p+q$. It computes the $\tau$-scale both for innovations reconstructed from the ARMA in \eqref{eq:arma-inno-recursion} and from the BIP-ARMA in \eqref{eq:bip-arma-model-inno-recursion}, and chooses as a final estimate $\hat{\boldsymbol{\beta}}_\tau^*$, which provides the smaller $\tau$-scale. We show, for iid innovations with a symmetric pdf, that the proposed estimator is strongly consistent with the ARMA model (Theorem 1 and Theorem 2). Further, the estimator is asymptotically normal for  the ARMA model case with a controllable efficiency with respect to the maximum-likelihood estimator (Theorem 3). Finally, an expression for the IF which measures robustness against infinitesimal contamination is provided.

\subsection{Definition of the $\tau$-estimator under the ARMA model}
Let $\hat{\sigma}_n^M(\boldsymbol{a}_n(\boldsymbol{\beta}))$ be an M-estimate of the scale of $\boldsymbol{a}_n(\boldsymbol{\beta})=(a_{p+1}(\boldsymbol{\beta}), \ldots, a_{n}(\boldsymbol{\beta}))$ based on $\rho_1(x)$ which satisfies {\bf A3}, i.e.,
\begin{equation}
 \frac{1}{n-p}\sum_{t=p+1}^n \rho_1\left(\frac{a_t(\boldsymbol{\beta})}{\hat{\sigma}_n^M(\boldsymbol{a}_n(\boldsymbol{\beta}))}\right)=b.
\end{equation}
{\bf (A5)} {\it Assume that $\mathrm{sup\ }\rho_1(x)>b$.} \\
The $\tau$-estimate of $\boldsymbol{\beta}=(\boldsymbol{\phi},\boldsymbol{\theta},\mu)$ under the ARMA model is defined according to
\begin{equation}
 \label{eq:tau_parameter_estimate}
 \hat{\boldsymbol{\beta}}_{\tau}= \mathrm{\text{arg }}\underset{\boldsymbol{\beta}\in\mathcal{B}}{\mathrm{\text{min }}} \hat{\sigma}_n^\tau(\boldsymbol{a}_n(\boldsymbol{\beta})),
\end{equation}
where $\hat{\sigma}_n^\tau(\boldsymbol{a}_n(\boldsymbol{\beta}))$ is the $\tau$-estimate \cite{yohai-1988} of the scale of $\boldsymbol{a}_n(\boldsymbol{\beta})$ and is defined as
\begin{equation}
\label{eq:tau-arma-inno-scale}
\hat{\sigma}_n^\tau(\boldsymbol{a}_n(\boldsymbol{\beta}))=\hat{\sigma}_n^M(\boldsymbol{a}_n(\boldsymbol{\beta}))\sqrt{\frac{1}{n-p}\sum_{t=p+1}^n\rho_2\left(\frac{a_t(\boldsymbol{\beta})}{\hat{\sigma}_n^M(\boldsymbol{a}_n(\boldsymbol{\beta}))}\right)}
\end{equation}
\if\paper\singlecol
Here $\mathcal{B}=\mathcal{B}_{0}\times \mathbb{R}$ where $\mathcal{B}_{0}= \left\{ (\boldsymbol{\phi},\boldsymbol{\theta})\in \mathbb{R}^{p+q}: |z|\geq 1 + \zeta \mathrm{\text{ holds for all roots }} z \mathrm{\text{ of }} \phi(B) \mathrm{\text{ and }} \theta(B) \right\}$ for some small $\zeta>0$. \\
\else
Here $\mathcal{B}=\mathcal{B}_{0}\times \mathbb{R}$ where $\mathcal{B}_{0}= \left\{ (\boldsymbol{\phi},\boldsymbol{\theta})\in \mathbb{R}^{p+q}: |z|\geq 1 + \zeta\right. \mathrm{\text{ holds}} $ $\left.\mathrm{\text{for all roots }} z \mathrm{\text{ of }} \phi(B)  \mathrm{\text{ and }} \theta(B) \right\}$ for some small $\zeta>0$. \\
\fi
{\bf (A6)} {\it Assume that $\rho_2(x)$ satisfies {\bf A3}, and additionally, $2\rho_2(x)-\psi_2(x)x\geq0$, where $\psi_2(x)=\frac{d\rho_2(x)}{dx}$.}

\subsection{Definition of the $\tau$-estimator under the BIP ARMA model}
The $\tau$-estimate of $\boldsymbol{\beta}=(\boldsymbol{\phi},\boldsymbol{\theta},\mu)$ under the BIP-ARMA model is defined according to
\begin{equation}
 \label{eq:tau_bip_parameter_estimate}
 \hat{\boldsymbol{\beta}}_{\tau}^b= \mathrm{\text{arg }}\underset{\boldsymbol{\beta}\in\mathcal{B}}{\mathrm{\text{min }}} \hat{\sigma}_n^\tau(\boldsymbol{a}_n^b(\boldsymbol{\beta},\hat{\sigma}(\boldsymbol{\beta}))),
\end{equation}
where 
\if\paper\singlecol
\begin{equation}
\label{eq:tau-bip-arma-inno-scale}
\hat{\sigma}_n^\tau(\boldsymbol{a}_n^b(\boldsymbol{\beta},\hat{\sigma}(\boldsymbol{\beta})))=\hat{\sigma}_n^M(\boldsymbol{a}_n^b(\boldsymbol{\beta},\hat{\sigma}(\boldsymbol{\beta})))\sqrt{\frac{1}{n-p}\sum_{t=p+1}^n\rho_2\left(\frac{a_t^b(\boldsymbol{\beta},\hat{\sigma}(\boldsymbol{\beta}))}{\hat{\sigma}_n^M(\boldsymbol{a}_n^b(\boldsymbol{\beta},\hat{\sigma}(\boldsymbol{\beta})))}\right)}
\end{equation}
\else
\begin{eqnarray}
\label{eq:tau-bip-arma-inno-scale}
\hat{\sigma}_n^\tau(\boldsymbol{a}_n^b(\boldsymbol{\beta},\hat{\sigma}(\boldsymbol{\beta})))&=&\hat{\sigma}_n^M(\boldsymbol{a}_n^b(\boldsymbol{\beta},\hat{\sigma}(\boldsymbol{\beta})))\cdot\nonumber  \\
& &\!\!\! \sqrt{\frac{1}{n-p}\sum_{t=p+1}^n\rho_2\left(\frac{a_t^b(\boldsymbol{\beta},\hat{\sigma}(\boldsymbol{\beta}))}{\hat{\sigma}_n^M(\boldsymbol{a}_n^b(\boldsymbol{\beta},\hat{\sigma}(\boldsymbol{\beta})))}\right)}\nonumber \\
\end{eqnarray}
\fi
and $\boldsymbol{a}_n^b(\boldsymbol{\beta},\hat{\sigma}(\boldsymbol{\beta}))=(a_{p+1}^b(\boldsymbol{\beta},\hat{\sigma}(\boldsymbol{\beta})), \ldots, a_{n}^b(\boldsymbol{\beta},\hat{\sigma}(\boldsymbol{\beta})))$ is recursively obtained from (\ref{eq:bip-arma-model-inno-recursion}). To compute $\hat{\sigma}(\boldsymbol{\beta})$, the MA-infinity representation of the BIP-ARMA model is used
\begin{equation}
 y_t = \mu - a_t + \sum_{i=1}^\infty \lambda_i\sigma\eta\left(\frac{a_{t-i}}{\sigma}\right),
\end{equation}
where $\lambda_i(\boldsymbol{\beta})$ are the coefficients of $\phi^{-1}(B)\theta(B)$. It then follows that
\begin{equation}
\label{eq:bip-ma-sigma}
\sigma^2(\boldsymbol{\beta})=\frac{\sigma_y^2}{1+\kappa^2\sum_{i=1}^{\infty}\lambda_i^2(\boldsymbol{\beta})},
\end{equation}
where $\sigma_y$ is the standard deviation of $y_t$ and
\begin{equation}
\kappa^2=\mathrm{Var}\left[\eta\left(\frac{a_t}{\sigma}\right)\right]=\mathrm{E}\left[\left( \eta\left(\frac{a_t}{\sigma}\right) - \mathrm{E}\left[\eta\left(\frac{a_t}{\sigma}\right)\right] \right)^2 \right].
\end{equation}
The estimate of $\sigma$ in Eq.~(\ref{eq:bip-ma-sigma}) can then be computed according to
\begin{equation}
\label{eq:bip-ma-sigma-est}
\hat{\sigma}^2(\boldsymbol{\beta})=\frac{\hat{\sigma}_n^\tau(\boldsymbol{y}_n)^2}{1+\kappa^2\sum_{i=1}^{q_{\mathrm{long}}}\lambda_i^2(\boldsymbol{\beta})},
\end{equation}
with $\boldsymbol{y}_n=(y_1,\ldots,y_n)$, and where $q_{\mathrm{long}}$ is chosen sufficiently large to approximate the MA-infinity representation.

\subsection{Definition of the proposed $\tau$-estimator}
The final $\tau$-estimate of the innovations scale is
\begin{equation}
\label{eq:final_inno_scale}
 \hat{\sigma}_{\tau}^*= \mathrm{min}\left\{\hat{\sigma}_n^\tau(\boldsymbol{a}_n(\hat{\boldsymbol{\beta}}_{\tau})),\hat{\sigma}_n^\tau(\boldsymbol{a}_n^b(\hat{\boldsymbol{\beta}}_{\tau}^b,\hat{\sigma}(\hat{\boldsymbol{\beta}}_{\tau}^b)))\right\}
\end{equation}
and the final parameter estimate becomes
\begin{equation}
\label{eq:final_beta_tau}
\hat{\boldsymbol{\beta}}_\tau^*=\left\{
                \begin{array}{lll}
                   \hat{\boldsymbol{\beta}}_{\tau} & \mathrm{if}\quad \hat{\sigma}_n^\tau(\boldsymbol{a}_n(\hat{\boldsymbol{\beta}}_{\tau}))<\hat{\sigma}_n^\tau(\boldsymbol{a}_n^b(\hat{\boldsymbol{\beta}}_{\tau}^b,\hat{\sigma}(\hat{\boldsymbol{\beta}}_{\tau}^b))) \\
                  \hat{\boldsymbol{\beta}}_{\tau}^b & \mathrm{if}\quad \hat{\sigma}_n^\tau(\boldsymbol{a}_n^b(\hat{\boldsymbol{\beta}}_{\tau}^b,\hat{\sigma}(\hat{\boldsymbol{\beta}}_{\tau}^b)))<\hat{\sigma}_n^\tau(\boldsymbol{a}_n(\hat{\boldsymbol{\beta}}_{\tau})). \\
                \end{array}
              \right.
\end{equation}
It is shown in Sec.~\ref{subsec:stat_anal} that when the data follows an ARMA model without outliers, the result that  $\hat{\sigma}_n^\tau(\boldsymbol{a}_n(\hat{\boldsymbol{\beta}}_{\tau}))<\hat{\sigma}_n^\tau(\boldsymbol{a}_n^b(\hat{\boldsymbol{\beta}}_{\tau}^b,\hat{\sigma}(\hat{\boldsymbol{\beta}}_{\tau}^b)))$ is asymptotically obtained for $n\rightarrow \infty$. This implies that the asymptotic efficiency of $\hat{\boldsymbol{\beta}}_\tau^*$ is independent of $\eta$. However, this does not hold in the finite sample size case. 

\subsection{Statistical analysis}
\label{subsec:stat_anal}
{\bf Theorem 1.} establishes strong consistency of the $\tau$-estimator of the ARMA parameters.\\
{\it Assume that $y_t$ follows from Eq. (\ref{eq:y_t_ARMA}) with $a_t$ satisfying {\bf A1}. Further, assume that $\rho_1$ satisfies {\bf A3} and {\bf A5} and that $\rho_2$ satisfies {\bf  A6}. Then, the $\tau$-estimator $\hat{\boldsymbol{\beta}}_{\tau}$ defined in Eq. (\ref{eq:tau_parameter_estimate}) is strongly consistent for $\boldsymbol{\beta}_0$.}\\
Proving this theorem requires Lemmas 1-3.\\
{\bf Lemma 1.} provides the Fisher consistency of the $\tau$-estimator of the ARMA parameters given all past observations\footnote{For visual clarity, let $\hat{\sigma}_\tau(a_t^e(\boldsymbol{\beta}_0))=:\hat{\sigma}_\tau(\boldsymbol{\beta}_0)$, $\hat{\sigma}_\tau(a_t^e(\boldsymbol{\beta}))=:\hat{\sigma}_\tau(\boldsymbol{\beta})$ and $\hat{\sigma}_M(a_t^e(\boldsymbol{\beta}))=:\hat{\sigma}_M(\boldsymbol{\beta})$}.\\
{\it Let $y_t$ be an observation from an ARMA(p,q), as in Eq.~(\ref{eq:y_t_ARMA}). Assume that $\rho_1(x)$ is bounded and satisfies {\bf A3} and {\bf A5}. It then holds, with $\sigma_0$ denoting the true innovations scale, that $\sigma_0=\hat{\sigma}_\tau(\boldsymbol{\beta}_0)<\hat{\sigma}_\tau(\boldsymbol{\beta})$ if $\boldsymbol{\beta}\in\mathcal{B}$ and $\boldsymbol{\beta}\neq \boldsymbol{\beta}_0$. This implies that the estimate $\hat{\boldsymbol{\beta}}_{\tau}$, as defined in Eq.~(\ref{eq:tau_parameter_estimate}), is Fisher consistent for $\boldsymbol{\beta}_0$.}
\begin{proof}
Consider the assumptions made in Theorem 1 which are the same assumptions made in Lemma 2 in \cite{Muler-2007}. This lemma states: if $\boldsymbol{\beta}\in\mathcal{B}$ and $\boldsymbol{\beta}\neq\boldsymbol{\beta}_0$ it holds, for an M-estimate of scale $\hat{\sigma}_M(\boldsymbol{\beta})>0$ defined by 
\begin{equation}
\label{eq:mscale_functional}
 \mathrm{E}\left[\rho_1\left(\frac{a_t^e(\boldsymbol{\beta})}{\hat{\sigma}^M(\boldsymbol{\beta})} \right)\right]=b,
\end{equation}
that $\hat{\sigma}_M(\boldsymbol{\beta}_0)<\hat{\sigma}_M(\boldsymbol{\beta})$. Since for $\boldsymbol{\beta}\neq \boldsymbol{\beta}_0$
\begin{equation}
a_t^e(\boldsymbol{\beta})=  \omega(B)a_t+c(\mu_0-\mu)
\end{equation}
where
\begin{equation}
 \omega(B)=\theta^{-1}(B)\theta_0(B)\phi_0^{-1}(B)\phi(B)=1+\sum_{i=1}^\infty \omega_iB^i
\end{equation}
and 
\begin{equation}
 c=\frac{1-\sum_{i=1}^p\phi_i}{1-\sum_{i=1}^q\theta_i}\neq0
\end{equation}
it follows by defining 
\begin{equation}
 \Delta_t(\boldsymbol{\beta})=\sum_{i=1}^\infty\omega_i a_{t-i}+c(\mu_0-\mu), 
\end{equation}
that
\begin{eqnarray}
\label{eq:tau_asymptotic}
 \hat{\sigma}_\tau^2(\boldsymbol{\beta})	&=&	\hat{\sigma}_M^2(\boldsymbol{\beta})\mathrm{E}\left[\rho_2\left(\frac{a_t^e(\boldsymbol{\beta})}{\hat{\sigma}_M(\boldsymbol{\beta})} \right)\right]\nonumber \\
				&=&	\hat{\sigma}_M^2(\boldsymbol{\beta})\mathrm{E}\left[\rho_2\left(\frac{a_t+\Delta_t(\boldsymbol{\beta})}{\hat{\sigma}_M(\boldsymbol{\beta})} \right)\right]
\end{eqnarray}
Using Lemma 3.1 (i) from \cite{tau-yohai1986-techrep} it then follows that
\begin{equation}
 \hat{\sigma}_\tau^2(\boldsymbol{\beta}) >	\hat{\sigma}_M^2(\boldsymbol{\beta})\mathrm{E}\left[\rho_2\left(\frac{a_t}{\hat{\sigma}_M(\boldsymbol{\beta})} \right)\right]
\end{equation}
for all $\Delta_t(\boldsymbol{\beta})\neq0$. Then, using Lemma 3.1 (ii) from \cite{tau-yohai1986-techrep}, and assuming that $\rho_2(x)$ is continuously differentiable, it is sufficient to show, for $\hat{\sigma}_M>0$, that
\begin{equation}
 h(\hat{\sigma}_M)=\hat{\sigma}_M^2\mathrm{E}\left[\rho_2\left(\frac{a_t}{\hat{\sigma}_M} \right)\right]
\end{equation}
is nondecreasing with respect to $\hat{\sigma}_M$, since {\bf A6} implies that
\begin{equation}
 \frac{dh(\hat{\sigma}_M)}{d\hat{\sigma}_M}=\hat{\sigma}_M\mathrm{E}\left[2\rho_2\left(\frac{a_t}{\hat{\sigma}_M}\right)-\psi_2\left(\frac{a_t}{\hat{\sigma}_M}\right)\frac{a_t}{\hat{\sigma}_M}\right]\geq0.
\end{equation}
\end{proof}
{\bf Lemma 2.} states the almost sure convergence of the $\tau$-estimator of the innovations scale to the population value based on the expectation operator.\\
{\it Under the assumptions of Theorem 1, for any $d>0$, it follows that
\begin{equation}
\label{eq:lemma1}
 \lim_{n\rightarrow \infty} \underset{\boldsymbol{\beta}\in\mathcal{B}_{0}\times [-d,d]}{\mathrm{sup\ }}| \hat{\sigma}_n^\tau(\boldsymbol{a}_n(\boldsymbol{\beta}))-  \hat{\sigma}_\tau(\boldsymbol{\beta})|=0\quad\mathrm{a.s.}.
\end{equation}
}
\begin{proof}
The continuity and positivity of the M-scale functional $\hat{\sigma}_M(\boldsymbol{\beta})>0$ defined in Eq.~(\ref{eq:mscale_functional}) was shown in Lemma 5 of \cite{Muler-2007}. The continuity and positivity of $\hat{\sigma}_\tau(\boldsymbol{\beta})$ follows from (\ref{eq:tau_asymptotic}), as long as $\rho_2$ satisfies {\bf A3}. Let 
\begin{equation}
 h_1 = \underset{\boldsymbol{\beta}\in\mathcal{B}_{0}\times [-d,d]}{\mathrm{inf}}\hat{\sigma}_\tau(\boldsymbol{\beta})
\end{equation}
and
\begin{equation}
 h_2 = \underset{\boldsymbol{\beta}\in\mathcal{B}_{0}\times [-d,d]}{\mathrm{sup\ }}\hat{\sigma}_\tau(\boldsymbol{\beta}).
\end{equation}
Then $h_1>0$ and $h_2<\infty$. 
According to Lemma 5 of \cite{Muler-2007}, it holds, for any $d>0$, that
\begin{equation}
\label{eq:muler07_lemma5}
 \lim_{n\rightarrow \infty} \underset{\boldsymbol{\beta}\in\mathcal{B}_{0}\times [-d,d]}{\mathrm{sup\ }}| \hat{\sigma}_n^M(\boldsymbol{a}_n(\boldsymbol{\beta}))-  \hat{\sigma}_M(\boldsymbol{\beta})|=0\quad\mathrm{a.s.}.
\end{equation}
From Lemma 2 of \cite{muler-2002}, it holds, under the assumptions {\bf A3}, {\bf A6}  on $\rho_j$, $j=1,2$, that 
\if\paper\singlecol
\begin{equation}
\label{eq:muler02_lemma2}
\lim_{n\rightarrow \infty} \underset{\boldsymbol{\beta}\in\mathcal{B}_{0}\times [-d,d],c\in[h_1/2,2h_2]}{\mathrm{sup\ }} \left| \frac{1}{n-p}\sum_{t=p+1}^n\rho_j\left(\frac{a_t^e(\boldsymbol{\beta})}{c}\right)-\mathrm{E}\left[\rho_j\left(\frac{a_t^e(\boldsymbol{\beta})}{c}\right) \right]  \right|=0\quad\mathrm{a.s.}.
\end{equation}
\else
\begin{eqnarray}
\label{eq:muler02_lemma2}
\lim_{n\rightarrow \infty} \underset{\boldsymbol{\beta}\in\mathcal{B}_{0}\times [-d,d],c\in[h_1/2,2h_2]}{\mathrm{sup\ }}& &\left| \frac{1}{n-p}\sum_{t=p+1}^n\rho_j\left(\frac{a_t^e(\boldsymbol{\beta})}{c}\right)\right. \nonumber \\
& & \left.-\mathrm{E}\left[\rho_j\left(\frac{a_t^e(\boldsymbol{\beta})}{c}\right) \right]  \right|=0\quad\mathrm{a.s.}.\nonumber \\
\end{eqnarray}
\fi
Eq.~(\ref{eq:lemma1}) then follows from (\ref{eq:muler07_lemma5}), (\ref{eq:muler02_lemma2}) and (\ref{eq:tau_asymptotic}).
\end{proof}
\noindent {\bf Lemma 3}\\
{\it Under the assumptions of Theorem 1, there exists $d>0$, such that
\begin{equation}
\label{eq:lemma2}
 \lim_{n\rightarrow \infty} \underset{|\mu|>d,(\boldsymbol{\phi},\boldsymbol{\theta})\in\mathcal{B}_{0}}{\mathrm{inf}} \quad \mathrm{inf}\quad \hat{\sigma}_n^\tau(\boldsymbol{a}_n(\boldsymbol{\beta})) > \sigma_0+1 \quad \mathrm{a.s.}.
\end{equation}
}\\
\begin{proof}
The proof is based on the one given in Lemma 6 of \cite{Muler-2007} which states that
 \begin{equation}
\label{eq:muler07_lemma6}
 \lim_{n\rightarrow \infty} \underset{|\mu|>d,(\boldsymbol\phi,\boldsymbol{\theta})\in\mathcal{B}_{0}}{\mathrm{inf}} \quad \mathrm{inf} \quad \hat{\sigma}_n^M(\boldsymbol{a}_n(\boldsymbol{\beta})) > \sigma_0+1 \quad \mathrm{a.s.},
\end{equation}
where $\rho_1$ has been replaced by $\rho_j$, $j=1,2$, and $\rho_j$ is assumed to be consistent with {\bf A3} and {\bf A6}. Then, using the continuity and positivity of $\hat{\sigma}_\tau(\boldsymbol{\beta})$ and the definition of the $\tau$-scale of (\ref{eq:tau_asymptotic}), (\ref{eq:lemma2}) follows from (\ref{eq:muler07_lemma6}).
\end{proof}
\begin{proof}[Proof of Theorem 1] 
Take $\xi>0$ arbitrarily small and let $d$ be as in Lemma 3. The continuity of the M-scale functional $\hat{\sigma}_M(\boldsymbol{\beta})>0$ defined in (\ref{eq:mscale_functional}) follows from Lebesgue's dominated convergence theorem. The continuity of $\hat{\sigma}_\tau(\boldsymbol{\beta})$ follows from (\ref{eq:tau_asymptotic}) as long as $\rho_2$ satisfies {\bf A3}. By Lemma 1 of this paper, there exists $0<\gamma<1$ such that
\begin{equation}
 \underset{\boldsymbol{\beta}\in\mathcal{B}_{0}\times [-d,d],||\boldsymbol{\beta}-\boldsymbol{\beta}_0||\geq \xi}{\mathrm{min}} \quad \hat{\sigma}_\tau(\boldsymbol{\beta})\geq \sigma_0+\gamma.
\end{equation}
By Lemma 2 of this paper, there exists $n_1$, such that for $n\geq n_1$
\begin{equation}
 \underset{\boldsymbol{\beta}\in\mathcal{B}_{0}\times [-d,d],||\boldsymbol{\beta}-\boldsymbol{\beta}_0||\geq \xi}{\mathrm{min}} \quad \hat{\sigma}_n^\tau(\boldsymbol{a}_n(\boldsymbol{\beta}))\geq \sigma_0+\gamma/2
\end{equation}
and
\begin{equation}
\hat{\sigma}_n^\tau(\boldsymbol{a}_n(\boldsymbol{\beta}_0))\leq \sigma_0 + \gamma/4.
\end{equation}
By Lemma 3, there exists $n_2$, such that for $n\geq n_2$
\begin{equation}
\underset{|\mu|>d,(\boldsymbol\phi,\boldsymbol{\theta})\in\mathcal{B}_{0}}{\mathrm{inf}}\quad \hat{\sigma}_n^\tau(\boldsymbol{a}_n(\boldsymbol{\beta}))> \sigma_0 + \gamma\quad \mathrm{a.s.}.
\end{equation}
Therefore, for $n\geq \max\{n_1,n_2\}$ it holds that $||\hat{\boldsymbol{\beta}}_{\tau}-\boldsymbol{\beta}_0||<\xi$, which proves the theorem.
\end{proof}
{\bf Theorem 2.} establishes, under an ARMA model, that the BIP $\tau$- is asymptotically equivalent to a $\tau$-estimator.\\
{\it Assume that $y_t$ follows (\ref{eq:y_t_ARMA}) with $a_t$ satisfying {\bf A1}. Further, assume that $\rho_1$ and $\rho_2$ are bounded, that $\rho_1$ satisfies {\bf A3} and {\bf A5}, that $\rho_2$ satisfies {\bf A6}, that $P(a_t\in \mathcal{C})<1$ for any compact set $\mathcal{C}$, and, finally, that $\eta$ satisfies {\bf A4}. Then, if $y_t$ is not white noise, with probability 1, there exists $n_0$, such that $\hat{\boldsymbol{\beta}}_{\tau}^b=\hat{\boldsymbol{\beta}}_{\tau}$ for all $n\geq n_0$ and then $\hat{\boldsymbol{\beta}}_{\tau}^*\rightarrow\boldsymbol{\beta}_0$ a.s..}
\begin{proof}
Theorem 2 of \cite{Muler-2007} shows that 
\begin{equation}
\label{eq:s_est_bip_equiv}
\underset{n\rightarrow \infty}{\mathrm{lim}}\underset{\boldsymbol{\beta}\in\mathcal{B}}{\mathrm{inf}}\ \hat{\sigma}_n^M(\boldsymbol{a}_n^b(\boldsymbol{\beta},\hat{\sigma}(\boldsymbol{\phi},\boldsymbol{\theta})))>\sigma_0+\delta \quad \mathrm{a.s.}.
\end{equation}
Starting from (\ref{eq:s_est_bip_equiv}),
\begin{equation}
\underset{n\rightarrow \infty}{\mathrm{lim}}\underset{\boldsymbol{\beta}\in\mathcal{B}}{\mathrm{inf}}\hat{\sigma}_n^\tau(\boldsymbol{a}_n^b(\boldsymbol{\beta},\hat{\sigma}(\boldsymbol{\phi},\boldsymbol{\theta})))>\sigma_0+\delta \quad \mathrm{a.s.}
\end{equation}
follows from Lemmas 9 and 10 of \cite{Muler-2007} together with Eq. (\ref{eq:tau_asymptotic}), as long as $\rho_2$ satisfies {\bf A3}. Furthermore, in Theorem 1, it is established that 
\begin{equation}
\underset{n\rightarrow \infty}{\mathrm{lim}} \hat{\sigma}_n^\tau(\boldsymbol{a}_n(\hat{\boldsymbol{\beta}}_\tau))=\sigma_0 \quad \mathrm{a.s.}.
\end{equation}
and this proves the theorem.
\end{proof}
{\bf Theorem 3.} establishes the asymptotic normality of the estimator for the ARMA model.\\
{\it Let $y_t$ be as in (\ref{eq:y_t_ARMA}), let {\bf A1}, {\bf A2}, {\bf A3} be fulfilled and let $\mathrm{E}[a_t^2]<\infty$. Further, assume that $\frac{d \psi_\tau(x)}{dx}$ and $\frac{d^2 \psi_\tau(x)}{dx^2}$ are continuous and bounded functions. Then, the $\tau$-estimator is asymptotically normally distributed with 
\begin{equation}
 (n-p)^{1/2}(\hat{\boldsymbol{\beta}}_{\tau}-\boldsymbol{\beta}_0)\xrightarrow[D]{}\mathcal{N}(\mathbf{0},\boldsymbol{\Sigma}),
\end{equation}
where 
\begin{equation}
\label{eq:distribution_of_BIPtau}
\boldsymbol{\Sigma}=\frac{\sigma_0^2\mathrm{E}\left[\psi_\tau^2(a_t/\sigma_0)\right]}{\mathrm{E}^2\left[\psi'_\tau(a_t/\sigma_0)\right]}\left( 
\begin{array}{ll}
 \sigma^2 \mathbf{C}^{-1} & \mathbf{0}\\
\mathbf{0} & c_0^{-2}
\end{array}
\right)
\end{equation}
with $\psi'_\tau(x) = \frac{d \psi_\tau(x)}{dx}$,
\begin{equation}
c_0=-\frac{1-\sum_{i=1}^p\phi_{0i}}{1-\sum_{i=1}^q\theta_{0i}}
\end{equation}
and $\mathbf{C}$ being the matrix of dimensions $(p+q+1) \times (p+q+1)$ with elements
\begin{eqnarray}
c_{i,j}&=&\sum_{k=0}^\infty \nu_k\nu_{k+j-i} \quad \mathrm{if}\quad i\leq j \leq p, \\
c_{p+i,p+j}&=& \sum_{k=0}^\infty \varpi_k \varpi_{k+j-i} \quad     \mathrm{if}\quad i\leq j \leq q,\\
c_{i,p+j} &=& -\sum_{k=0}^\infty \varpi_k \nu_{k+j-i} \quad \mathrm{if} \quad i\leq p, j\leq p, i\leq j,\\
c_{i,p+j} &=& -\sum_{k=0}^\infty v_k \varpi_{k+i-j} \quad \mathrm{if} \quad i\leq p, j\leq q, j\leq i.              
\end{eqnarray}
}
Here $\phi_0^{-1}(B)=1+\sum_{i=1}^\infty \nu_iB^i$ and $\theta_0^{-1}(B)=1+\sum_{i=1}^\infty \varpi_i B^i$.
\begin{proof}
According to Theorem 5 of \cite{Muler-2007}, an M-estimator, under the same assumptions that are made in this theorem, is asymptotically normally distributed with 
\begin{equation}
 (n-p)^{1/2}(\hat{\boldsymbol{\beta}}_{M}-\boldsymbol{\beta}_0)\xrightarrow[D]{}\mathcal{N}(\mathbf{0},\boldsymbol{\Sigma}),
\end{equation}
where
\begin{equation}
\boldsymbol{\Sigma}=\frac{\sigma_0^2\mathrm{E}\left[\psi^2(a_t/\sigma_0)\right]}{\mathrm{E}^2\left[\psi'(a_t/\sigma_0)\right]}\left( 
\begin{array}{ll}
 \sigma^2 \mathbf{C}^{-1} & \mathbf{0}\\
\mathbf{0} & c_0^{-2}.
\end{array}
\right)
\end{equation}
To prove Theorem 3, it must be shown that the $\tau$-estimator of the ARMA parameters satisfies an $M$-estimating equation. Differentiating (\ref{eq:tau_parameter_estimate}) yields the following system of equations:
\if\paper\singlecol
\begin{eqnarray}
\label{eq:tau_equations}
\nabla \hat{\sigma}_n^\tau(\boldsymbol{a}_n(\boldsymbol{\beta}))^2&=&2\hat{\sigma}_n^M(\boldsymbol{a}_n(\boldsymbol{\beta}))\nabla \hat{\sigma}_n^M(\boldsymbol{a}_n(\boldsymbol{\beta}))\frac{1}{n-p}\sum_{t=p+1}^n\rho_2\left(\frac{a_t(\boldsymbol{\beta})}{\hat{\sigma}_n^M(\boldsymbol{a}_n(\boldsymbol{\beta}))}\right)\cdot\nonumber \\
& & +\frac{1}{n-p}\sum_{t=p+1}^n\psi_2\left(\frac{a_t(\boldsymbol{\beta})}{\hat{\sigma}_n^M(\boldsymbol{a}_n(\boldsymbol{\beta}))}\right)
\bigg( \nabla a_t(\boldsymbol{\beta}) \hat{\sigma}_n^M(\boldsymbol{a}_n(\boldsymbol{\beta}))-a_t(\boldsymbol{\beta})\nabla \hat{\sigma}_n^M(\boldsymbol{a}_n(\boldsymbol{\beta})) \bigg)=\boldsymbol{0}.\nonumber \\
\end{eqnarray}
\else
\begin{eqnarray}
\label{eq:tau_equations}
\nabla \hat{\sigma}_n^\tau(\boldsymbol{a}_n(\boldsymbol{\beta}))^2 \!\!\!\!  &=& \!\! 2\hat{\sigma}_n^M\!(\boldsymbol{a}_n(\boldsymbol{\beta}))\nabla \hat{\sigma}_n^M(\boldsymbol{a}_n(\boldsymbol{\beta}))\cdot \nonumber \\
\!\!& & \frac{1}{n-p}\sum_{t=p+1}^n\rho_2\left(\frac{a_t(\boldsymbol{\beta})}{\hat{\sigma}_n^M(\boldsymbol{a}_n(\boldsymbol{\beta}))}\right)\nonumber \\
\!\!& & +\frac{1}{n-p}\sum_{t=p+1}^n\psi_2\left(\frac{a_t(\boldsymbol{\beta})}{\hat{\sigma}_n^M(\boldsymbol{a}_n(\boldsymbol{\beta}))}\right)\cdot \nonumber \\
\!\!& &\! \!  \bigg(\! \nabla a_t(\boldsymbol{\beta}) \hat{\sigma}_n^M\!(\boldsymbol{a}_n(\boldsymbol{\beta}))\! -\! a_t(\boldsymbol{\beta})\nabla \hat{\sigma}_n^M\!(\boldsymbol{a}_n(\boldsymbol{\beta}))\! \bigg)\nonumber\!\!\!\!\!\!\!\! \\
\!\!& & =\boldsymbol{0}.
\end{eqnarray}
\fi
Here,
\begin{equation}
\label{eq:s_scale_derivative}
\nabla \hat{\sigma}_n^M(\boldsymbol{a}_n(\boldsymbol{\beta}))\! = \! -\hat{\sigma}_n^M(\!\boldsymbol{a}_n(\boldsymbol{\beta})\!)\frac{\sum_{t=p+1}^n\psi_1\left(\frac{a_t(\boldsymbol{\beta})}{\hat{\sigma}_n^M(\boldsymbol{a}_n(\boldsymbol{\beta}))}\right)\nabla a_t(\boldsymbol{\beta})}{\sum_{t=p+1}^n\psi_1\left(\frac{a_t(\boldsymbol{\beta})}{\hat{\sigma}_n^M(\boldsymbol{a}_n(\boldsymbol{\beta}))}\right) a_t(\boldsymbol{\beta})}
\end{equation}
with $\nabla a_t(\boldsymbol{\beta}) = \left(\frac{\partial a_t^e(\boldsymbol{\beta})}{\partial \phi_i},  \frac{\partial a_t^e(\boldsymbol{\beta})}{\partial \theta_j},  \frac{\partial a_t^e(\boldsymbol{\beta})}{\partial \mu}\right)^{\tp}$, where
\begin{equation}
 \frac{\partial a_t^e(\boldsymbol{\beta})}{\partial \phi_i} = -\theta^{-1}(B)(y_{t-i}-\mu), \quad 1 \leq i \leq p,
\end{equation}
\begin{equation}
 \frac{\partial a_t^e(\boldsymbol{\beta})}{\partial \theta_j} = -\theta^{-2}(B)\phi(B)(y_{t-j}-\mu), \quad 1 \leq j \leq q,
\end{equation}
and
\begin{equation}
 \frac{\partial a_t^e(\boldsymbol{\beta})}{\partial \mu} = - \frac{1-\sum_{i=1}^p\phi_i}{1-\sum_{j=1}^q\theta_j}.
\end{equation}
Replacing (\ref{eq:s_scale_derivative}) in (\ref{eq:tau_equations}) and defining
\if\paper\singlecol
\begin{equation}
 W_n(\boldsymbol{\beta}) = \frac{\sum_{t=p+1}^n 2 \rho_2\left(\frac{a_t(\boldsymbol{\beta})}{\hat{\sigma}_n^M(\boldsymbol{a}_n(\boldsymbol{\beta}))}\right)-\psi_2\left(\frac{a_t(\boldsymbol{\beta})}{\hat{\sigma}_n^M(\boldsymbol{a}_n(\boldsymbol{\beta}))}\right)\frac{a_t(\boldsymbol{\beta})}{\hat{\sigma}_n^M(\boldsymbol{a}_n(\boldsymbol{\beta}))}}{\sum_{t=p+1}^n \psi_1\left(\frac{a_t(\boldsymbol{\beta})}{\hat{\sigma}_n^M(\boldsymbol{a}_n(\boldsymbol{\beta}))}\right)\frac{a_t(\boldsymbol{\beta})}{\hat{\sigma}_n^M(\boldsymbol{a}_n(\boldsymbol{\beta}))}},
\end{equation}
\else
\begin{equation}
 W_n(\boldsymbol{\beta})\! = \!\frac{\!\!\sum_{t=p+1}^n 2 \rho_2\!\left(\frac{a_t(\boldsymbol{\beta})}{\hat{\sigma}_n^M(\boldsymbol{a}_n(\boldsymbol{\beta}))}\right)\!-\!\psi_2\!\left(\frac{a_t(\boldsymbol{\beta})}{\hat{\sigma}_n^M(\boldsymbol{a}_n(\boldsymbol{\beta}))}\!\right)\!\frac{a_t(\boldsymbol{\beta})}{\hat{\sigma}_n^M(\boldsymbol{a}_n(\boldsymbol{\beta}))}}{\sum_{t=p+1}^n \psi_1\left(\frac{a_t(\boldsymbol{\beta})}{\hat{\sigma}_n^M(\boldsymbol{a}_n(\boldsymbol{\beta}))}\right)\frac{a_t(\boldsymbol{\beta})}{\hat{\sigma}_n^M(\boldsymbol{a}_n(\boldsymbol{\beta}))}},
\end{equation}
\fi
if $\rho_2(x)$ satisfies {\bf A6}, the $\tau$-estimate satisfies an M-estimating equation
\begin{equation}
\sum_{t=p+1}^n  \psi_{\tau}\left(\frac{a_t(\boldsymbol{\beta})}{\hat{\sigma}_n^M(\boldsymbol{a}_n(\boldsymbol{\beta}))}\right)\nabla a_t(\boldsymbol{\beta}) = \boldsymbol{0}
\end{equation}
with data adaptive $\psi_{\tau}$ given by
\begin{equation}
\label{eq:psi-tau}
 \psi_{\tau}(x) =  W_n(\boldsymbol{\beta})\psi_1\left(\frac{a_t(\boldsymbol{\beta})}{\hat{\sigma}_n^M(\boldsymbol{a}_n(\boldsymbol{\beta}))}\right)+\psi_2\left(\frac{a_t(\boldsymbol{\beta})}{\hat{\sigma}_n^M(\boldsymbol{a}_n(\boldsymbol{\beta}))}\right).
\end{equation}
Special cases are (i) $\rho_2(x)=1/2x^2$ which results in $W_n(\boldsymbol{\beta})=0$ and the $\tau$-estimator being equivalent to an LS estimator, (ii) $\rho_1(x)=\rho_2(x)$ which results in the $\tau$-estimator being equivalent to an S-estimator. 
The asymptotic value of the estimator is defined by 
\begin{equation}
\label{eq:asymptotic-tau}
\lim_{n \rightarrow \infty} \sum_{t=p+1}^n  \psi_{\tau}\left(\frac{a_t(\boldsymbol{\beta})}{\hat{\sigma}_n^M(\boldsymbol{a}_n(\boldsymbol{\beta}))}\right)\nabla a_t(\boldsymbol{\beta}) = \boldsymbol{0}
\end{equation}
and under suitable regularity conditions, i.e., ergodicity, the interchange of limits is justified (e.g. by dominated convergence) to yield
\if\paper\singlecol
\begin{equation}
\label{eq:equival_asymp-tau}
\lim_{n \rightarrow \infty} \sum_{t=p+1}^n  \psi_{\tau}\left(\! \frac{a_t(\boldsymbol{\beta})}{\hat{\sigma}_n^M(\boldsymbol{a}_n(\boldsymbol{\beta}))}\right)\nabla \boldsymbol{a}_n(\boldsymbol{\beta}) = \mathrm{E}\left[  \psi_{\tau}\left(\frac{a_t(\boldsymbol{\beta})}{\hat{\sigma}_M(\boldsymbol{\beta})}\right)\nabla a_t(\boldsymbol{\beta})     \right].
\end{equation}
\else
\begin{equation}
\label{eq:equival_asymp-tau}
\lim_{n \rightarrow \infty}\! \sum_{t=p+1}^n\!\!\!\!  \psi_{\tau}\!\left(\! \frac{a_t(\boldsymbol{\beta})}{\hat{\sigma}_n^M(\boldsymbol{a}_n(\boldsymbol{\beta}))}\! \right)\!\nabla \boldsymbol{a}_n(\boldsymbol{\beta})\! =\! \mathrm{E}\!\left[\!  \psi_{\tau}\!\left(\!\frac{a_t(\boldsymbol{\beta})}{\hat{\sigma}_M(\boldsymbol{\beta})}\!\right)\!\nabla a_t(\boldsymbol{\beta})     \!\right].
\end{equation}
\fi
\end{proof}
From \eqref{eq:distribution_of_BIPtau}, it follows, for the outlier free ARMA model, where the innovations follow the standard Gaussian distribution $F$, that the statistical efficiency of our proposed estimator is given by:\\
\begin{equation*}
\mathrm{EFF}(\psi_\tau,F) =\frac{\sigma_0^2\mathrm{E}_F\left[\psi_\tau^2(a_t/\sigma_0)\right]}{\sigma^2\mathrm{E}_F^2\left[\psi'_\tau(a_t/\sigma_0)\right]}.
\end{equation*} 

\subsection{Influence function (IF) analysis}
To analyze the infinitesimal effect of outliers on the asymptotic estimate, the IF is computed. Assume that the observations follow an ARMA model that is contaminated by additive or replacement outliers as in (\ref{eq:contaminated_arma_ao}) or (\ref{eq:contaminated_arma_ro}). The temporal structure of the outliers may be patchy or iid, depending on the choice of the process $\xi_t^\varepsilon$. The dependent data IF is defined \cite{martin-1986} as the directional derivative at $F(x)$, i.e.,
\if\paper\singlecol
\begin{equation}
\label{eq:IF-Martin1986}
  \mathrm{IF}(\{F(x,\xi^\varepsilon,w)\};\hat{\boldsymbol{\beta}}_\infty)= \underset{\downarrow \varepsilon}{\lim}(\hat{\boldsymbol{\beta}}_\infty(F(y^\varepsilon))-\hat{\boldsymbol{\beta}}_\infty(F(x)))=\frac{\partial}{\partial \varepsilon}\hat{\boldsymbol{\beta}}_\infty(F(y^\varepsilon))|_{\varepsilon=0},
\end{equation}
\else
\begin{eqnarray}
\label{eq:IF-Martin1986}
  \mathrm{IF}(\{F(x,\xi^\varepsilon,w)\};\hat{\boldsymbol{\beta}}_\infty)&=& \underset{\downarrow \varepsilon}{\lim}(\hat{\boldsymbol{\beta}}_\infty(F(y^\varepsilon))-\hat{\boldsymbol{\beta}}_\infty(F(x)))\nonumber \\
&=& \frac{\partial}{\partial \varepsilon}\hat{\boldsymbol{\beta}}_\infty(F(y^\varepsilon))|_{\varepsilon=0},
\end{eqnarray}
\fi
provided that the limit exists. Here, $F(x)$, $F(w)$, $F(\xi^\varepsilon)$ and $F(y^\varepsilon)$ are the cdfs of $x_t$, $w_t$, $\xi^\varepsilon$ and $y_t^\varepsilon$, respectively. Further, $F(x,\xi^\varepsilon,w)$ is the joint distribution of $x_t$, $w_t$, $\xi^\varepsilon$. $\mathrm{IF}(\{F(x,\xi^\varepsilon,w)\};\hat{\boldsymbol{\beta}}_\infty)$ is defined for functionals which may be computed as a solution of the estimating equation
\begin{equation}
\label{eq:psi-type-estimator}
 \int \tilde{\psi}(\boldsymbol{y}_t,\hat{\boldsymbol{\beta}})dF(\boldsymbol{y}_t)=0.
\end{equation}
This class is quite large and contains both classical and robust parameter estimators, e.g. the M-estimators, the generalized M-estimators and estimators based on residual autocovariances (RA-estimators) \cite{martin-1986}. It will be shown that the $\tau$-estimators of the ARMA parameters are of the $\tilde{\psi}$-type.
\begin{proof}
From (\ref{eq:equival_asymp-tau}) it follows, by defining 
\begin{equation}
  \tilde{\psi}(\boldsymbol{y}_t,\hat{\boldsymbol{\beta}}_\tau)=\psi_{\tau}\left(\frac{a_t(\boldsymbol{\beta})}{\hat{\sigma}_M(\boldsymbol{\beta})}\right)\nabla \boldsymbol{a}_t(\boldsymbol{\beta}) 
\end{equation}
and by noting the results stated in (\ref{eq:psi-type-estimator}), that
\begin{equation}
\label{eq:tau-psi-estimator}
   \int  \psi_{\tau}\left(\frac{a_t(\boldsymbol{\beta})}{\hat{\sigma}_M(\boldsymbol{\beta})}\right)\nabla a_t(\boldsymbol{\beta}) dF(\boldsymbol{y}_t)= \int \tilde{\psi}(\boldsymbol{y}_t,\hat{\boldsymbol{\beta}}_\tau)dF(\boldsymbol{y}_t)=0.
\end{equation}
This proves that the $\tau$-estimator is a $\tilde{\psi}$-estimator. 
\end{proof}
\subsection*{IF of the $\tau$-estimator for an AR(1) with AO contamination}
In general, the IF defined by Eq.~(\ref{eq:IF-Martin1986}) is a curve on measure space. It is useful to compute the IF of the $\tau$-estimator for the particular case of AR(1) models with additive outliers\footnote{To the best of our knowledge, all IFs that have been explicitly computed in the literature concern AR(1) and MA(1) models only.}. 

Let $y^\varepsilon_t$ follow (\ref{eq:contaminated_arma_ao}) with $x_t$ satisfying (\ref{eq:y_t_ARMA}) with $p=1$, $q=0$ and $\mu=0$. Further, let the  $\xi_t^\varepsilon$ be an independently distributed 0-1 sequence that is independent of $x_t$ and $w_t$. Then, as long as the following assumptions are fulfilled:\\
\noindent{\bf (A7)}  {\it $\psi_{\tau}(\cdot)$ is continuous, odd, bounded, and $\psi_{\tau}(\infty)=0$,}\\
{\bf (A8)}  {\it $\frac{d\psi_{\tau}(x)}{dx}$ is bounded,}\\
{\bf (A9)}  {\it $|a_2(\phi_1) \psi_{\tau}(a_1(\phi_1))|\leq K|a_2(\phi_1)|$, with $K<\infty$,}\\
{\bf (A10)}  {\it $\frac{\partial a_2(\phi_1) \psi_{\tau}(a_1(\phi_1)))}{\partial a_1(\phi_1)}$, $\frac{\partial a_2(\phi_1) \psi_{\tau}(a_1(\phi_1)))}{\partial a_2(\phi_1)}$ are continuous,} \\
\if\paper\singlecol
{\bf (A11)}  {\it $\left|\frac{\partial (a_2(\phi_1) \psi_{\tau}(a_1(\phi_1))))}{\partial a_1(\phi_1)}\right|\leq K |a_2(\phi_1)|\quad and \quad \left|\frac{\partial (a_1(\phi_1) \psi_{\tau}(a_1(\phi_1))))}{\partial a_2(\phi_1)}\right|\leq K,$ with $K<\infty$,}\\
\else
{\bf (A11)}  {\it $\left|\frac{\partial (a_2(\phi_1) \psi_{\tau}(a_1(\phi_1))))}{\partial a_1(\phi_1)}\right|\leq K|a_2(\phi_1)|\quad and \quad $} {$\left|\frac{\partial (a_1(\phi_1) \psi_{\tau}(a_1(\phi_1))))}{\partial a_2(\phi_1)}\right|\leq K,$ with $K<\infty$,}\\
\fi
{\bf (A12)}  {\it $\mathrm{E}\left[|w_1|<\infty\right]$,}\\
the IF of the $\tau$-estimator is given by
\if\paper\singlecol
\begin{equation}
  \mathrm{IF}(F(w),\hat{\boldsymbol{\beta}}_\tau,\phi) = \frac{(1-\phi_1^2)^{1/2}}{\mathcal{E}_0}\mathrm{E}\left[(x_0+w_0)(1-\phi_1^2)^{1/2} \psi_\tau\left(a_1-\phi_1 w_0\right) \right] 
\end{equation}
\else
\begin{eqnarray}
  \mathrm{IF}(F(w),\hat{\boldsymbol{\beta}}_\tau,\phi)\!\! &=&\!\! \frac{(1-\phi_1^2)^{1/2}}{\mathcal{E}_0}\cdot \nonumber \\
& &\mathrm{E}\left[\!(x_0+w_0)(1-\phi_1^2)^{1/2} \psi_\tau\left(a_1\!-\!\phi_1 w_0\right)\! \right]\!\!\! \nonumber\\
\end{eqnarray}
\fi
Here $\mathcal{E}_0=E\left[\nu^2 \left.\frac{\partial (\psi_{\tau}\left(x\right))}{\partial x}\right|_{x=u} \right]\neq0$, where $\nu$ and $u$ are independent standard normal random variables.\\
\begin{proof}
With Theorem 1 and (\ref{eq:tau-psi-estimator}), as long as {\bf A7-A12}, hold, the proof follows the steps of Theorem 5.2 in \cite{martin-1986}, with $\psi(x)$ replaced by $\psi_{\tau}(x)$.
\end{proof}
If we now let $P(w_t=c_w)=1$ for a constant $c_w$, the IF has the appealing heuristic interpretation of displaying the influence of a contamination value $c_w$ on the estimator, similarly to Hampel's definition \cite{hampel-1974} for iid data. The computation of the IF then requires the evaluation of the following integrals:
\begin{eqnarray}
\label{eq:if-tau-ar1-ex0}
\mathcal{E}_0 = \int_{-\infty}^{\infty} \nu^2 \left.\frac{\partial (\psi_{\tau}\left(x\right))}{\partial x}\right|_{x=u} \frac{1}{2\pi}e^{-\frac{u^2+\nu^2}{2}}du d\nu
\end{eqnarray}
\if\paper\singlecol
\begin{eqnarray}
\label{eq:if-tau-ar1-ex1}
\!\!\!\!\!\!\mathcal{E}_1\!\!\!\! &=&\!\!\!\!  \int_{-\infty}^{\infty} \int_{-\infty}^{\infty} \int_{-\infty}^{\infty} (x_0+w_0)(1-\phi_1^2)^{1/2} \psi_\tau\left(a_1-\phi_1 w_0\right) f(x_1,x_0;\phi_1)f(w_0)dx_1 dx_0 dw_0
\end{eqnarray}
\else
\begin{eqnarray}
\label{eq:if-tau-ar1-ex1}
\!\!\!\!\!\!\mathcal{E}_1\!\!\!\! &=&\!\!\!\!  \int_{-\infty}^{\infty} \int_{-\infty}^{\infty} \int_{-\infty}^{\infty} (x_0+w_0)(1-\phi_1^2)^{1/2} \psi_\tau\left(a_1-\phi_1 w_0\right)\cdot \nonumber \\
& & f(x_1,x_0;\phi_1)f(w_0)dx_1 dx_0 dw_0
\end{eqnarray}
\fi
Here the following equality holds
\begin{equation}
f(x_1,x_0;\phi_1)=f(x_1|x_0;\phi_1)f(x_0;\phi_1)
\end{equation}
where
\begin{equation}
 f(x_1|x_0;\phi_1)=\frac{1}{\sqrt{2\pi} \sigma} e^{-\frac{1}{2}\frac{(x_1-\phi_1 x_0)^2}{\sigma^2}}
\end{equation}
\begin{equation}
 f(x_0;\phi_1)=\frac{\sqrt{1-\phi_1^2}}{\sqrt{2\pi} \sigma} e^{-\frac{1}{2}\frac{x_0^2(1-\phi_1^2)}{\sigma^2}}.
\end{equation}

Fig.~\ref{fig:dependent-bip-tau-if-ar1} displays the IF of the proposed estimator and that of the LS estimator for the above example of an AR(1) with $\phi=-0.5$ for independent AOs of magnitude $c_w$ for 
\if\paper\singlecol
\begin{equation}
 \label{eq:tau-rho}
 \rho_2(x)=\left\{\begin{array}{ll}
                            0.5x^2 & \mathrm{\text{if }} |x|\leq 2\\
                            0.002x^8 - 0.052x^6 + 0.432x^4-0.972x^2+1.792 & \mathrm{\text{if }} 2<|x|\leq 3\\
			    3.25 & |x|> 3,\\
                           \end{array}\right.
\end{equation}
\else
\begin{equation}
 \label{eq:tau-rho}
 \rho_2(x)=\left\{\begin{array}{ll}
                            0.5x^2 & \mathrm{\text{if }} |x|\leq 2\\
                            0.002x^8 - 0.052x^6  & \\
			    + 0.432x^4-0.972x^2+1.792 & \mathrm{\text{if }} 2<|x|\leq 3\\
			    3.25 & |x|> 3,\\
                           \end{array}\right.
\end{equation}
\fi
$\rho_1(x)=\rho_2(x/c_1)$, with $c_1 = 0.4050$ and $\eta(x)=d\rho_2(x)/dx$. By comparing this figure to Fig.~1 in \cite{martin-1986}, we conclude that the gross-error sensitivity (GES), which is defined as the supremum of $|\mathrm{IF}(F(w),\hat{\boldsymbol{\beta}},\phi)|$ of our estimator is smaller than that of the generalized M-estimator (GM) and the residual autocovariance (RA) estimator. The comparison with Fig.~4.2 of \cite{chakhchoukh-2010t}, leads to the deduction that the GES of our estimator is also smaller than that of the 
median-of-ratios-estimator (MRE) and ratio-of-medians-estimator (RME), which were published in \cite{chakhchoukh-2010t,chakhchoukh-2010tsp}.

\if\paper\singlecol
\begin{figure}[htp]
   \centering
     \includegraphics[width=0.55\textwidth]{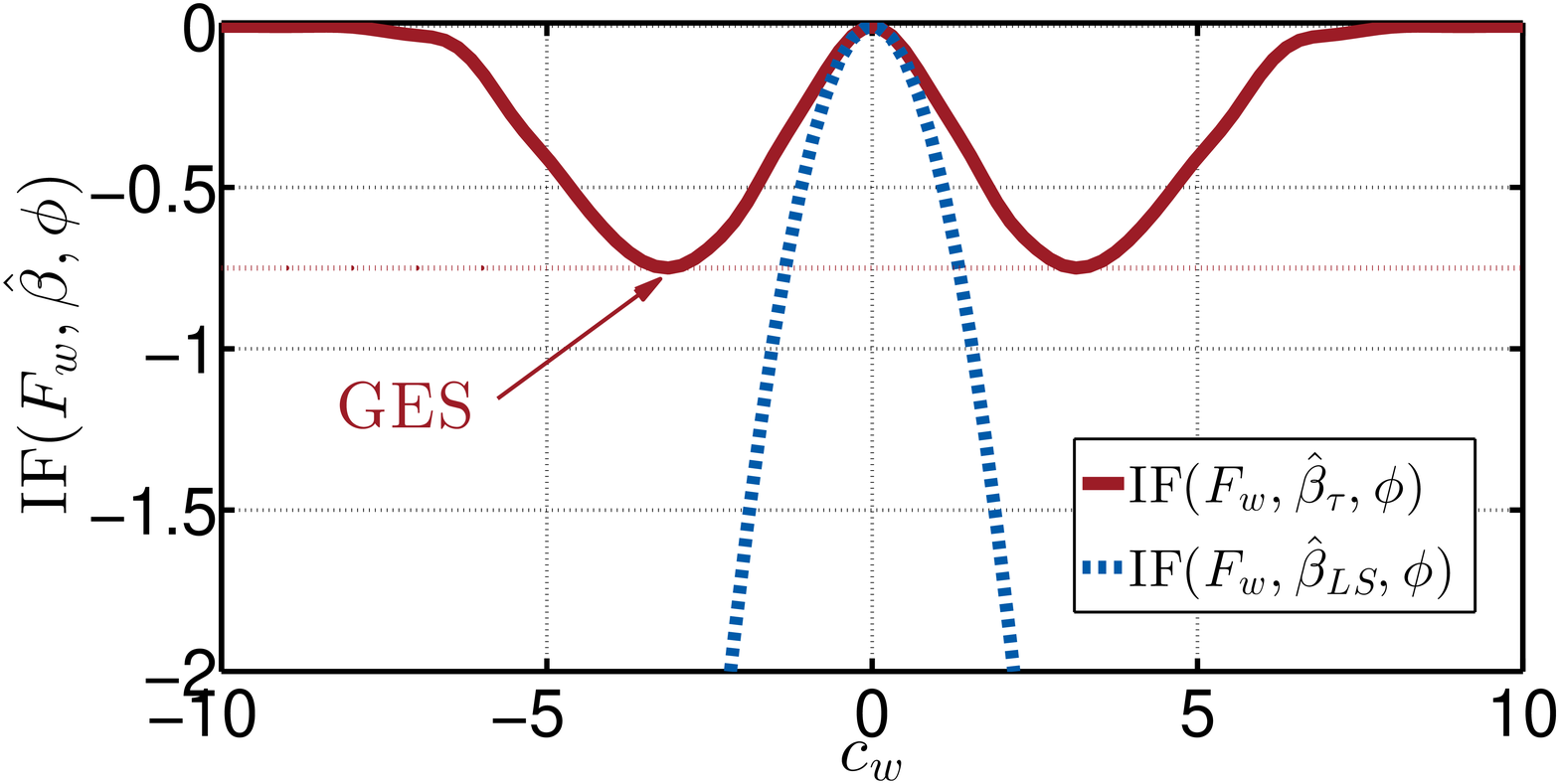} 
     \caption{The IF of the proposed estimator and that of the LS estimator for the AR(1) model with $\phi=-0.5$ and for the case of independent AOs of magnitude $c_w$. The supremum of the IF is the gross-error sensitivity (GES).}
     \label{fig:dependent-bip-tau-if-ar1}
 \end{figure}
\else
\begin{figure}[htp]
   \centering
     \includegraphics[width=0.5\textwidth]{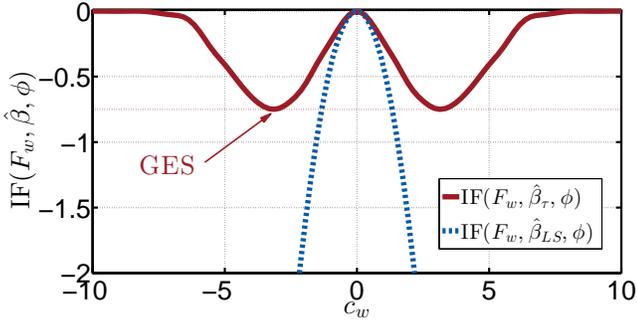} 
      \vspace{-25 pt}
     \caption{The IF of the proposed estimator and that of the LS estimator for the AR(1) model with $\phi=-0.5$ and for the case of independent AOs of magnitude $c_w$. The supremum of the IF is the gross-error sensitivity (GES).}
     \label{fig:dependent-bip-tau-if-ar1}
 \end{figure}
\fi

\vspace{-5 pt}
\section{Algorithm}
\label{sec:algorithm}
\subsection{Estimating the AR parameters with a Robust Durbin-Levinson Algorithm}
\label{subsec:ar-algorithms}
To compute $\hat{\boldsymbol{\beta}}_\tau^*$ for the AR($p$) model, a robust Durbin-Levinson type algorithm is proposed, where the parameters are recursively found for $m=1,\ldots,p$. 
Table~\ref{table:pseudocodeAR1} details the algorithm for the AR(1) model, while Fig.~\ref{fig:dependent-bip-tau-algo-ar1} illustrates the procedure by giving an example\footnote{First evaluating (\ref{eq:tau-arma-inno-scale}), (\ref{eq:tau-bip-arma-inno-scale}) on a coarse grid (e.g. using a step size of $\Delta_{\zeta^0}=0.05$) and then modeling the true curves by a polynomial is an optional step to speed up the algorithm compared to evaluating (\ref{eq:tau-arma-inno-scale}), (\ref{eq:tau-bip-arma-inno-scale}) on a very fine grid. Details are given in Sec. \ref{subsec:complexity}.}.
\begin{table}[htp]
\footnotesize
\begin{center}
 \begin{tabular}{l}
\toprule
{\bf{Algorithm 1: Robust Durbin-Levinson Algorithm for the AR(1)}} \\
\midrule
{{\bf for} $p=1$, $q=0$, $\zeta^0=-0.99:\Delta_{\zeta^0}:0.99$}\\
\phantom{bla} { {\bf compute} AR(1) innovations from (\ref{eq:arma-inno-recursion}) and (\ref{eq:bip-arma-model-inno-recursion})} \\
\phantom{blabla}  $\rightarrow\boldsymbol{a}_n(\zeta^0), \boldsymbol{a}_n^b(\zeta^0,\hat{\sigma}(\zeta^0))$ \\
\phantom{bla}  { {\bf compute} $\tau$-scale from (\ref{eq:tau-arma-inno-scale}), (\ref{eq:tau-bip-arma-inno-scale}) { with $\hat{\sigma}_n^{M}$}}\\
\phantom{blabla}  { computed as in \cite{maronna-2006} pages 40-41}\\
\phantom{blabla}  $\rightarrow\hat{\sigma}_{\tau}(\boldsymbol{a}_n(\zeta^0))$, and $\hat{\sigma}_{\tau}(\boldsymbol{a}_n^b(\zeta^0,\hat{\sigma}(\zeta^0)))$ \\
{{\bf end for}} \\
\phantom{bla}  {\bf fit polynomial to}  \\
\phantom{blabla} ($\zeta^0$,$\hat{\sigma}_{\tau}(\boldsymbol{a}_n(\zeta^0))$), and ($\zeta^0$,$\hat{\sigma}_{\tau}(\boldsymbol{a}_n^b(\zeta^0,\hat{\sigma}(\zeta^0)))$)\\
\phantom{blabla} at $\zeta^0=-0.99:\Delta_{\zeta^0}:0.99$\\ 
\phantom{bla} {\bf Estimate AR(1) by}  \\
\phantom{blabla} $\hat{\phi}_1 = \underset{\zeta}{\mathrm{argmin}} \ \left\{ \hat{\sigma}_{\tau}(\boldsymbol{a}_n(\zeta)),\hat{\sigma}_{\tau}(\boldsymbol{a}_n^b(\zeta,\hat{\sigma}(\zeta)))\right\}.$\\
\bottomrule
\end{tabular}
\end{center}
\caption{Summary of the robust Durbin-Levinson for the AR(1) model.}
\label{table:pseudocodeAR1}
\end{table}
The top graph depicts the results for $y_t=x_t$ with $\phi_1=-0.5$ for $\sigma=1$, $n=1000$. The bottom graph displays an illustrative AO example, where $\xi_t^\varepsilon w_t$ in (\ref{eq:contaminated_arma_ao}) produces 10 \% equally spaced AOs of amplitude 10.

For a general AR($p$) process, the parameters are found recursively for $m=2,\ldots,p$ by minimizing
\begin{equation}
\label{eq:bip-tau-algo-arm}
\hat{\phi}_{m,m} = \underset{\zeta}{\mathrm{argmin}} \ \left\{ \hat{\sigma}_{\tau}(\boldsymbol{a}_n(\zeta)),\hat{\sigma}_{\tau}(\boldsymbol{a}_n^b(\zeta,\hat{\sigma}(\zeta)))\right\}
\end{equation}
at each order $m$ in the same manner described in Table~\ref{table:pseudocodeAR1}, with the help of the Durbin-Levinson recursion:
\begin{equation}
 \label{eq:durbin-levinson}
   \hat{\phi}_{m,m} =\begin{cases} \zeta & \quad \mathrm{\text{if}} \quad i=m \\ \hat{\phi}_{m-1,i}-\zeta \hat{\phi}_{m-1,m-i} & \quad \mathrm{\text{if}} \quad 1\leq i \leq m-1 \end{cases}
\end{equation}

\if\paper\singlecol
\begin{figure}[htp]
   \centering
     \includegraphics[width=0.55\textwidth]{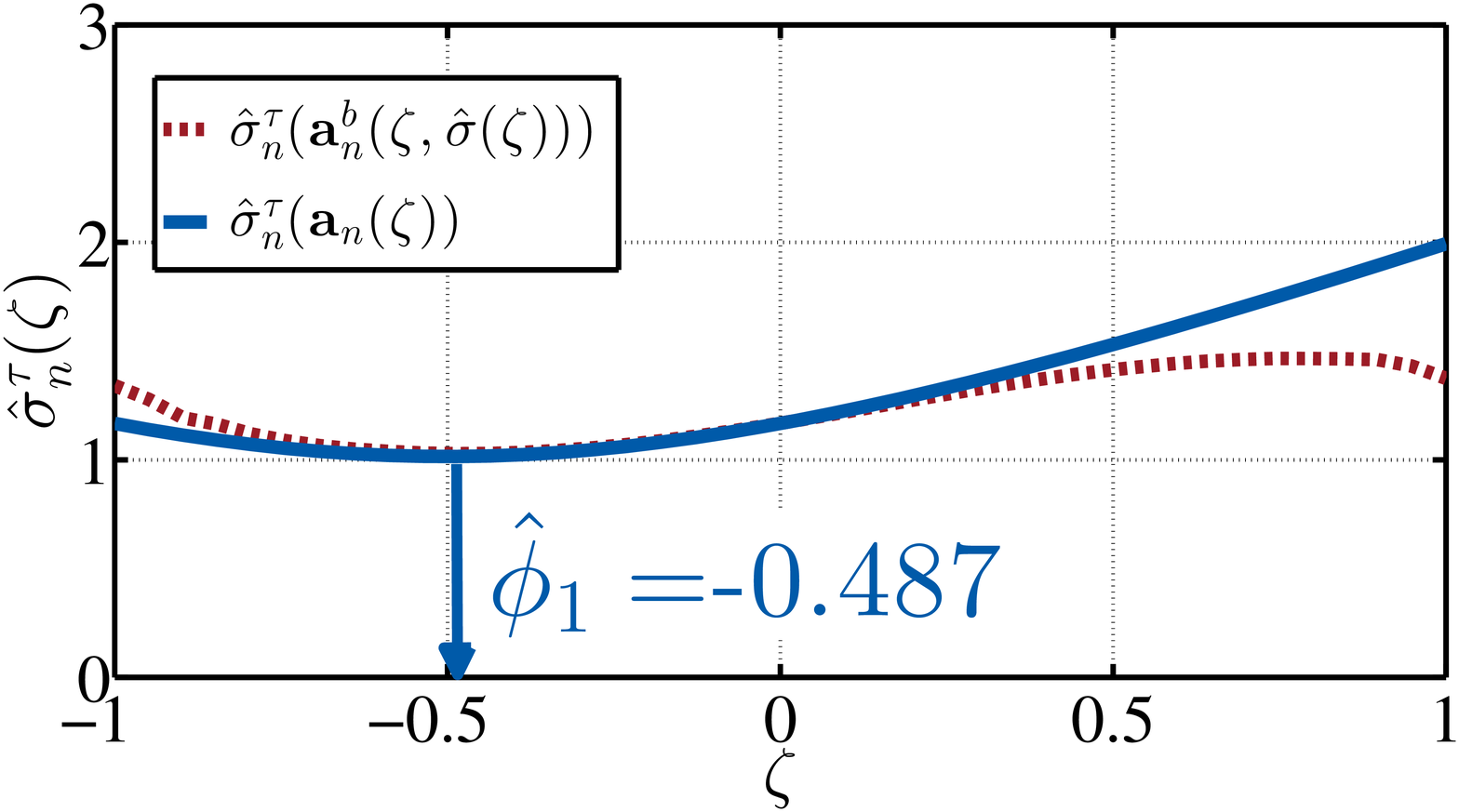} \\ 
     \includegraphics[width=0.55\textwidth]{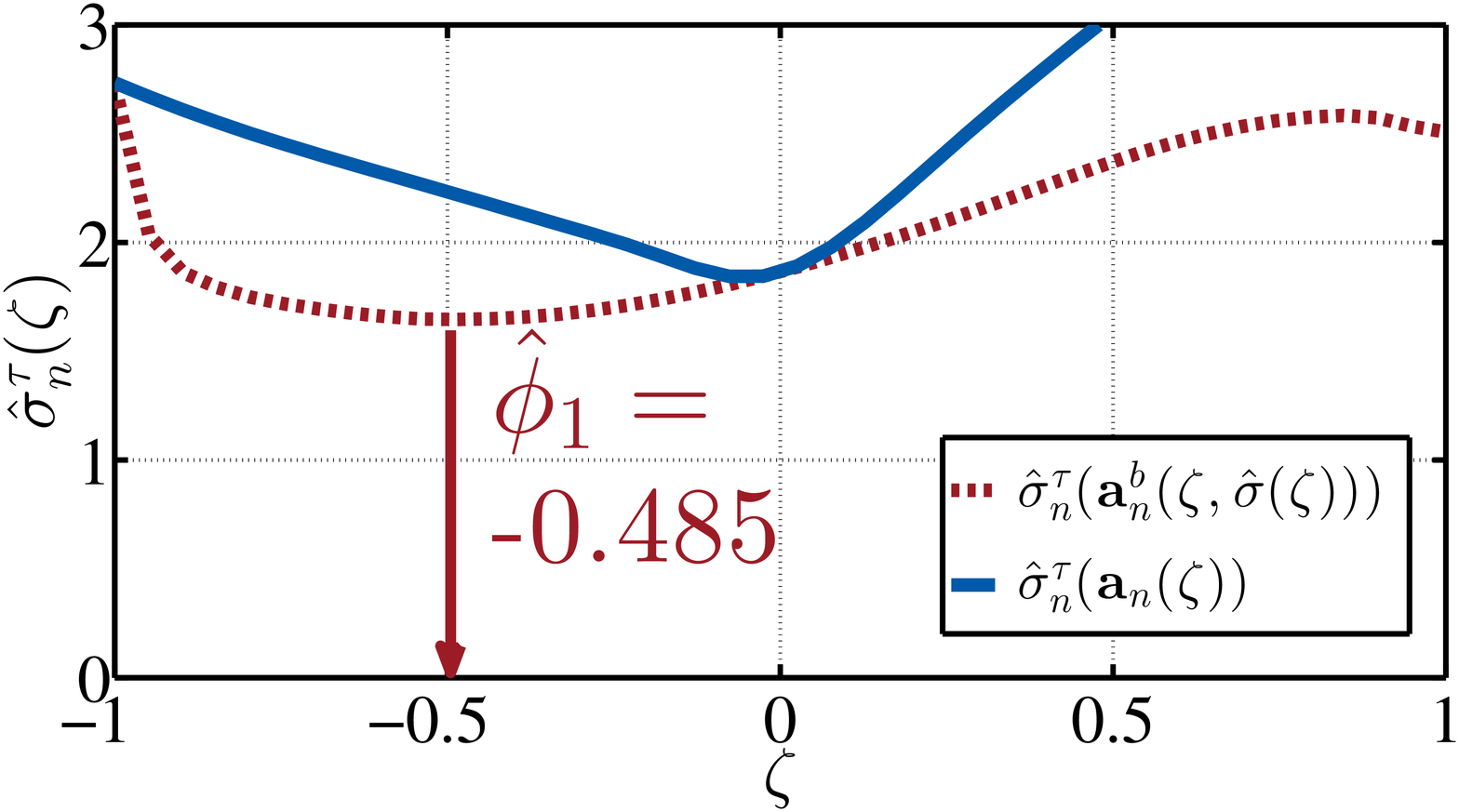} 
     \caption{Example of finding $-1<\zeta<1$ which minimizes  $\hat{\sigma}_n^\tau(\boldsymbol{a}_n(\boldsymbol{\zeta}))$and $\hat{\sigma}_n^\tau(\boldsymbol{a}_n^b(\boldsymbol{\zeta},\hat{\sigma}(\boldsymbol{\zeta})))$ for an AR(1) process with $\phi_1=-0.5$ and $\sigma=1$. (top) $y_t=x_t$ clean data example; (bottom) 10 \% equally spaced AOs of amplitude 10.}
     \label{fig:dependent-bip-tau-algo-ar1}
 \end{figure}
\else
\begin{figure}[htp]
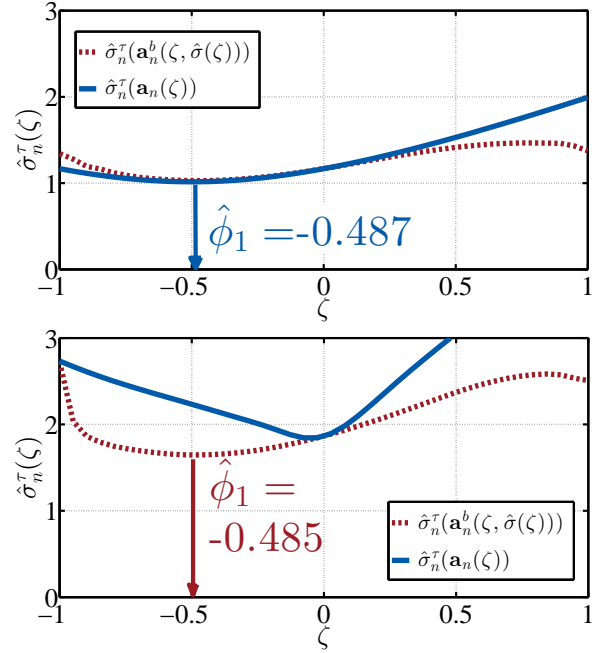

   \centering
     \includegraphics[width=0.5\textwidth]{fig3.eps} \\ 
     \includegraphics[width=0.5\textwidth]{fig3b.eps} 
      \vspace{-25 pt}
     \caption{Example of finding $-1<\zeta<1$ which minimizes  $\hat{\sigma}_n^\tau(\boldsymbol{a}_n(\boldsymbol{\zeta}))$and $\hat{\sigma}_n^\tau(\boldsymbol{a}_n^b(\boldsymbol{\zeta},\hat{\sigma}(\boldsymbol{\zeta})))$ for an AR(1) process with $\phi_1=-0.5$ and $\sigma=1$. (top) $y_t=x_t$ clean data example; (bottom) 10 \% equally spaced AOs of amplitude 10.}
     \label{fig:dependent-bip-tau-algo-ar1}
 \end{figure}
\fi 

\subsection{Estimating the AR parameters with a Robust Forward-Backward Algorithm}
\label{subsec:ar-algorithms-2}
In classical AR estimation, it is well known \cite{stoica2005spectral} that algorithms, which are based on forward and backward innovations estimates, outperform the Durbin-Levinson method. The following algorithm adapts the concept of minimizing the arithmetic mean of the forward and backward innovations estimates of scale.

The backward innovations estimates under the AR model are recursively obtained for $t=p+1,\ldots,n$ by:
\begin{equation}
\label{eq:ar-inno-backward-recursion}
a_t^{e,\mathrm{bw}}(\boldsymbol{\beta})=  y_{t-p} - \mu - \sum_{i=1}^p \phi_i(y_{t-p+i}-\mu)
\end{equation}
Similarly, for the BIP-AR model the backward innovations estimates are defined recursively for $n-p-1,\ldots,p+1$ by:
\if\paper\singlecol
\begin{equation}
\label{eq:bip-ar-inno-backward-recursion}
a_t^{b,\mathrm{bw}}(\boldsymbol{\beta},\sigma)=  y_{t-p} - \mu - \sum_{i=1}^p\phi_i(y_{t-p+i}-\mu) +\sum_{i=1}^p\bigg(\phi_i a_{t+i}^{b,\mathrm{bw}}(\boldsymbol{\beta},\sigma)-\phi_i\sigma \eta\bigg(\frac{a_{t+i}^{b,\mathrm{bw}}(\boldsymbol{\beta},\sigma)}{\sigma} \bigg) \bigg)
\end{equation}
\else
\begin{eqnarray}
\label{eq:bip-ar-inno-backward-recursion}
a_t^{b,\mathrm{bw}}(\boldsymbol{\beta},\sigma)\!&=&\!  y_{t-p} - \mu - \sum_{i=1}^p\phi_i(y_{t-p+i}-\mu)\nonumber\quad \\
\!& &\! +\sum_{i=1}^p\bigg(\phi_i a_{t+i}^{b,\mathrm{bw}}(\boldsymbol{\beta},\sigma)-\phi_i\sigma \eta\bigg(\frac{a_{t+i}^{b,\mathrm{bw}}(\boldsymbol{\beta},\sigma)}{\sigma} \bigg) \bigg) \quad \quad
\end{eqnarray}
\fi
\if\paper\singlecol
The $\tau$-scales of $\boldsymbol{a}^{\mathrm{bk}}_n(\boldsymbol{\beta})\!=\!(a^{\mathrm{bk}}_{p+1}(\boldsymbol{\beta}), \ldots, a^{\mathrm{bk}}_{n}(\boldsymbol{\beta}))$, $\boldsymbol{a}^{b,\mathrm{bk}}_n(\boldsymbol{\beta},\hat{\sigma}(\boldsymbol{\beta}))\!=\!(a^{b,\mathrm{bk}}_{p+1}(\boldsymbol{\beta},\hat{\sigma}(\boldsymbol{\beta})), \ldots, a^{b,\mathrm{bk}}_{n-p-1}(\boldsymbol{\beta},\hat{\sigma}(\boldsymbol{\beta})))$ are computed analogously to \eqref{eq:tau-arma-inno-scale} and \eqref{eq:tau-bip-arma-inno-scale} with $\hat{\sigma}(\boldsymbol{\beta})$ as given in \eqref{eq:bip-ma-sigma-est}.
\else
The $\tau$-scale estimtes of $\boldsymbol{a}^{\mathrm{bk}}_n(\boldsymbol{\beta})\!=\!(a^{\mathrm{bk}}_{p+1}(\boldsymbol{\beta}), \ldots, a^{\mathrm{bk}}_{n}(\boldsymbol{\beta}))$ and $\boldsymbol{a}^{b,\mathrm{bk}}_n(\boldsymbol{\beta},\hat{\sigma}(\boldsymbol{\beta}))\!=\!(a^{b,\mathrm{bk}}_{p+1}(\boldsymbol{\beta},\hat{\sigma}(\boldsymbol{\beta})), \ldots, a^{b,\mathrm{bk}}_{n-p-1}(\boldsymbol{\beta},\hat{\sigma}(\boldsymbol{\beta})))$ are computed analogously to \eqref{eq:tau-arma-inno-scale} and \eqref{eq:tau-bip-arma-inno-scale} with $\hat{\sigma}(\boldsymbol{\beta})$ as given in \eqref{eq:bip-ma-sigma-est}.
\fi

Table~\ref{table:pseudocodeAR1_fb} details the forward-backward algorithm for the AR(1) model. For a general AR($p$) models, the parameters are found recursively for $m=2,\ldots,p$ by means of \eqref{eq:durbin-levinson} and evaluation of
\if\paper\singlecol
\begin{eqnarray}
\label{eq:bip-tau-algo-arm}
\hat{\phi}_{m,m} &=& 
\underset{\zeta}{\mathrm{argmin}} \ \bigg\{(\hat{\sigma}_{\tau}(\boldsymbol{a}_n(\zeta))+\hat{\sigma}_{\tau}(\boldsymbol{a}^{\mathrm{bk}}_n(\zeta)))/2,
(\hat{\sigma}_{\tau}(\boldsymbol{a}_n^b(\zeta))+\hat{\sigma}_{\tau}(\boldsymbol{a}^{b,\mathrm{bk}}_n(\zeta)))/2,\big\}
\end{eqnarray}
\else
\begin{eqnarray}
\label{eq:bip-tau-algo-arm}
\!\!\!\!\hat{\phi}_{m,m}\!\!&=& \!\!
\underset{\zeta}{\mathrm{argmin}} \ \big\{(\hat{\sigma}_{\tau}(\boldsymbol{a}_n(\zeta))+\hat{\sigma}_{\tau}(\boldsymbol{a}^{\mathrm{bk}}_n(\zeta)))/2,\nonumber \\
\!\!& & \!\!\qquad\qquad(\hat{\sigma}_{\tau}(\boldsymbol{a}_n^b(\zeta))+\hat{\sigma}_{\tau}(\boldsymbol{a}^{b,\mathrm{bk}}_n(\zeta)))/2,\big\}
\end{eqnarray}
\fi
for each $m$ analogously to the AR(1) model case.

\begin{table}[htp]
\footnotesize
\begin{center}
 \begin{tabular}{l}
\toprule
{\bf{Algorithm 2: Robust Forward-Backward Algorithm for the AR(1)}} \\
\midrule
{{\bf for} $p=1$, $q=0$, $\zeta^0=-0.99:\Delta_{\zeta^0}:0.99$}\\
\phantom{bla} { {\bf compute} AR(1) innovations from (\ref{eq:arma-inno-recursion}),(\ref{eq:bip-arma-model-inno-recursion}),(\ref{eq:ar-inno-backward-recursion}) and (\ref{eq:bip-ar-inno-backward-recursion})} \\
\phantom{blabla}  $\rightarrow \boldsymbol{a}_n(\zeta^0), \boldsymbol{a}_n^b(\zeta^0,\hat{\sigma}(\zeta^0))$, $\boldsymbol{a}^{\mathrm{bk}}_n(\zeta^0), \boldsymbol{a}_n^{b,\mathrm{bk}}(\zeta^0,\hat{\sigma}(\zeta^0))$  \\
\phantom{bla}  { {\bf compute} $\tau$-scale from (\ref{eq:tau-arma-inno-scale}), (\ref{eq:tau-bip-arma-inno-scale}) { with $\hat{\sigma}_n^{M}$}}\\
\phantom{blabla}  { computed as in \cite{maronna-2006} pages 40-41}\\
\phantom{blabla}  $\rightarrow\hat{\sigma}_{\tau}(\boldsymbol{a}_n(\zeta^0))$, $\hat{\sigma}_{\tau}(\boldsymbol{a}_n^b(\zeta^0,\hat{\sigma}(\zeta^0)))$ \\
\phantom{blabla}  $\rightarrow\hat{\sigma}_{\tau}(\boldsymbol{a}^{\mathrm{bk}}_n(\zeta^0))$, $\hat{\sigma}_{\tau}(\boldsymbol{a}_n^{b,\mathrm{bk}}(\zeta^0,\hat{\sigma}(\zeta^0)))$ \\
{{\bf end for}} \\
\phantom{bla}  {\bf fit polynomial to}  \\
\phantom{blabla} ($\zeta^0$,$\hat{\sigma}_{\tau}(\boldsymbol{a}_n(\zeta^0))$), ($\zeta^0$,$\hat{\sigma}_{\tau}(\boldsymbol{a}_n^b(\zeta^0,\hat{\sigma}(\zeta^0)))$),\\
\phantom{blabla} ($\zeta^0$,$\hat{\sigma}_{\tau}(\boldsymbol{a}^{\mathrm{bk}}_n(\zeta^0))$), ($\zeta^0$,$\hat{\sigma}_{\tau}(\boldsymbol{a}_n^{b,\mathrm{bk}}(\zeta^0,\hat{\sigma}(\zeta^0)))$)\\
\phantom{blabla} at $\zeta^0=-0.99:\Delta_{\zeta^0}:0.99$\\ 
\phantom{bla} {\bf Estimate AR(1) by}  \\
\phantom{blabla} $\hat{\phi}_1 = \underset{\zeta}{\mathrm{argmin}} \ \left\{(\hat{\sigma}_{\tau}(\boldsymbol{a}_n(\zeta))+\hat{\sigma}_{\tau}(\boldsymbol{a}^{\mathrm{bk}}_n(\zeta)))/2,\right.$\\
\phantom{blabla} \qquad\qquad\qquad\qquad\qquad $(\hat{\sigma}_{\tau}(\boldsymbol{a}_n^b(\zeta))+\hat{\sigma}_{\tau}(\boldsymbol{a}^{b,\mathrm{bk}}_n(\zeta)))/2,\big\}$\\
\bottomrule
\end{tabular}
\end{center}
\caption{Summary of the robust forward-backward algorithm for the AR(1) model.}
\label{table:pseudocodeAR1_fb}
\end{table}

\subsection{Estimating the ARMA parameters}
Determining an estimate for $\boldsymbol{\beta}$ with $q>0$ requires finding the $\boldsymbol{\beta}$ that minimizes \eqref{eq:tau-arma-inno-scale} and \eqref{eq:tau-bip-arma-inno-scale}. Since this is a non-convex problem, the crucial point is to find a starting point that is sufficiently close to the true $\boldsymbol{\beta}$. Due to the computational complexity, except for some very simple cases (e.g. $p+q\leq2$), it is not possible to perform an exhaustive grid search. The following procedure to find a robust starting point is therefore proposed. 

\subsubsection{Robust starting point algorithm}
From \eqref{eq:bip-arma-model} it follows, for the AR model, that the one step prediction of $y_t$ can be computed recursively for $t\geq p+1$ via:
\if\paper\singlecol
\begin{equation}
\label{eq:bip-arma-model-prediction}
\hat{y}_t = \mu + \sum_{i=1}^p\phi_i \bigg(y_{t-i}-\mu- a_{t-i}^b(\hat{\boldsymbol{\beta}},\hat{\sigma})+\hat{\sigma} \eta\left(\frac{a_{t-i}^b(\hat{\boldsymbol{\beta}},\hat{\sigma})}{\hat{\sigma}} \right)\bigg).
\end{equation}
\else
\begin{eqnarray}
\label{eq:bip-arma-model-prediction}
\!\!\!\!\!\hat{y}_t\!\!\!\!\!\! &=&\!\!\!\!\!\! \mu + \sum_{i=1}^p\phi_i \bigg(y_{t-i}-\mu- a_{t-i}^b(\hat{\boldsymbol{\beta}},\hat{\sigma})+\hat{\sigma} \eta\left(\frac{a_{t-i}^b(\hat{\boldsymbol{\beta}},\hat{\sigma})}{\hat{\sigma}} \right)\bigg)\quad
\end{eqnarray}
\fi
From \eqref{eq:bip-arma-model-prediction}, outlier-cleaned observations are obtained for $t\geq p+1$ by computing
\begin{equation}
\label{eq:bip-cleaning}
y_t^* = y_t - a_t^b(\hat{\boldsymbol{\beta}},\hat{\sigma}) + \hat{\sigma} \eta\left(\frac{a_{t
}^b(\hat{\boldsymbol{\beta}},\hat{\sigma})}{\hat{\sigma}} \right).
\end{equation}
To find a starting point for the ARMA parameter estimation, the data is first cleaned from outliers using an AR($p$) approximation, which can be computed with the methods described in Sec. \ref{subsec:ar-algorithms} and \ref{subsec:ar-algorithms-2}. The choice of $p$ to be used in the approximation is discussed in Sec. \ref{subsec:ar_order_approx}. The starting point $\hat{\boldsymbol{\beta}}_0$ for the BIP-$\tau$ ARMA parameter estimation algorithm can then be computed, based on $y_t^*$, and by using any classical ARMA parameter estimator, e.g. \cite{jones-1980}.

\subsubsection{ARMA parameter estimation algorithm}
From \eqref{eq:tau-arma-inno-scale} and \eqref{eq:tau-bip-arma-inno-scale}, it is evident that the minimization of $(\hat{\sigma}_n^\tau(\boldsymbol{a}_n(\hat{\boldsymbol{\beta}}_{\tau})))^2$ and $(\hat{\sigma}_n^\tau(\boldsymbol{a}_n^b(\hat{\boldsymbol{\beta}}_{\tau}^b,\hat{\sigma}(\hat{\boldsymbol{\beta}}_{\tau}^b))))^2$ can be solved by any nonlinear LS algorithm, e.g. the Marquard algorithm. The initialization $\hat{\boldsymbol{\beta}}_0$, which is critical for the success of the Marquard algorithm is found via the robust starting point algorithm that is described above\footnote{We would like to highlight that the ARMA parameter estimation is performed on the original data $y_t$ and AR approximation based outlier cleaning is only used within the starting point algorithm to find $\hat{\boldsymbol{\beta}}_0$.}.

\vspace{-5 pt}
\section{Numerical Experiments}
\label{sec:simulations}
\subsection{Quantile bias curve analysis}
The maximum bias curve (MBC) provides information on the maximum asymptotic bias of an estimator w.r.t. a given fraction of contamination $\varepsilon$. For dependent data, the MBC is defined as for the iid case, but also depends on the outlier model. In practice, in the dependent data setting, the MBC is usually approximated by using Monte Carlo simulations \cite{maronna-2006,chakhchoukh-2009c,chakhchoukh-2010j} according to
\begin{equation}
\label{eq:dependent-mbc}
 \mathrm{MBC}(\varepsilon)=\underset{c_w}{\mathrm{sup }}\left|\hat{\boldsymbol{\beta}}_n(\varepsilon,c_w)-\boldsymbol{\beta}\right|
\end{equation}
The approximation is done by choosing, for $\mathrm{MBC}(\varepsilon)$, the worst-case estimate of $\boldsymbol{\beta}$ over all Monte Carlo runs for a given contamination probability $\varepsilon$. $c_w$ is a deterministic value that is varied on a grid such that for each value of $c_w$, the distribution of $w_t$ (see \eqref{eq:contaminated_arma_ro}) is given by $\mathrm{Pr}(w_t=-c_w)=\mathrm{Pr}(w_t=c_w)=0.5$. 
%

More generally, let
\begin{equation}
\label{eq:q-bias-ar1}
\mathrm{\text{QBC}\alpha(\varepsilon)}=Q_{\alpha} \left\{\left|\hat{\boldsymbol{\beta}}_n(\varepsilon,c_w)-\boldsymbol{\beta} \right|\right\}.
\end{equation}
denote the {\it quantile bias curve}, which states that $\alpha$ percent of the sorted data is to the left of $Q_{\alpha}$. For example, $\mathrm{\text{QBC}75(\varepsilon)}$ represents the MBC obtained in 75 \% of the Monte Carlo runs for varying $c_w$ and fixed $\varepsilon$. $\mathrm{\text{QBC}50(\varepsilon)}$  corresponds to the $\mathrm{\text{Median BC}(\varepsilon)}$ and $\mathrm{\text{QBC}100(\varepsilon)}$ is the $\mathrm{MBC(\varepsilon)}$. 

The quantile bias curves of the BIP $\tau$-estimator for the AR(1) model with independent AOs are provided in the bottom graph of Fig.~\ref{fig:dependent-bip-tau-qbc-ar1}. The top graph shows the maximum bias for a given pair of $(c_w,\varepsilon)$. As in \cite{muler-2009}, $\phi=0.5$ and the asymptotic value was approximated using $n=10000$. It can be seen from  Fig.~\ref{fig:dependent-bip-tau-qbc-ar1} (bottom) that the MBC saturates at 0.5 for $\varepsilon\geq 0.38$. This breakdown, however, only occurs for a 
minority of the data, as can be seen from the $\mathrm{\text{QBC}\alpha(\varepsilon)}$ with $\alpha<100$. Similar to the BIP-MM-estimator of \cite{muler-2009}, it is observed that the bias curves re-descend. This is easily explained by the fact that for large values of $\varepsilon$ the probability of obtaining patches of outliers increases. The effect of the patches is to increase the correlation, and therewith, to prevent a further shrinkage of the estimates towards zero. 
\if\paper\singlecol
\begin{figure}[htp]
   \centering
     \includegraphics[width=0.55\textwidth]{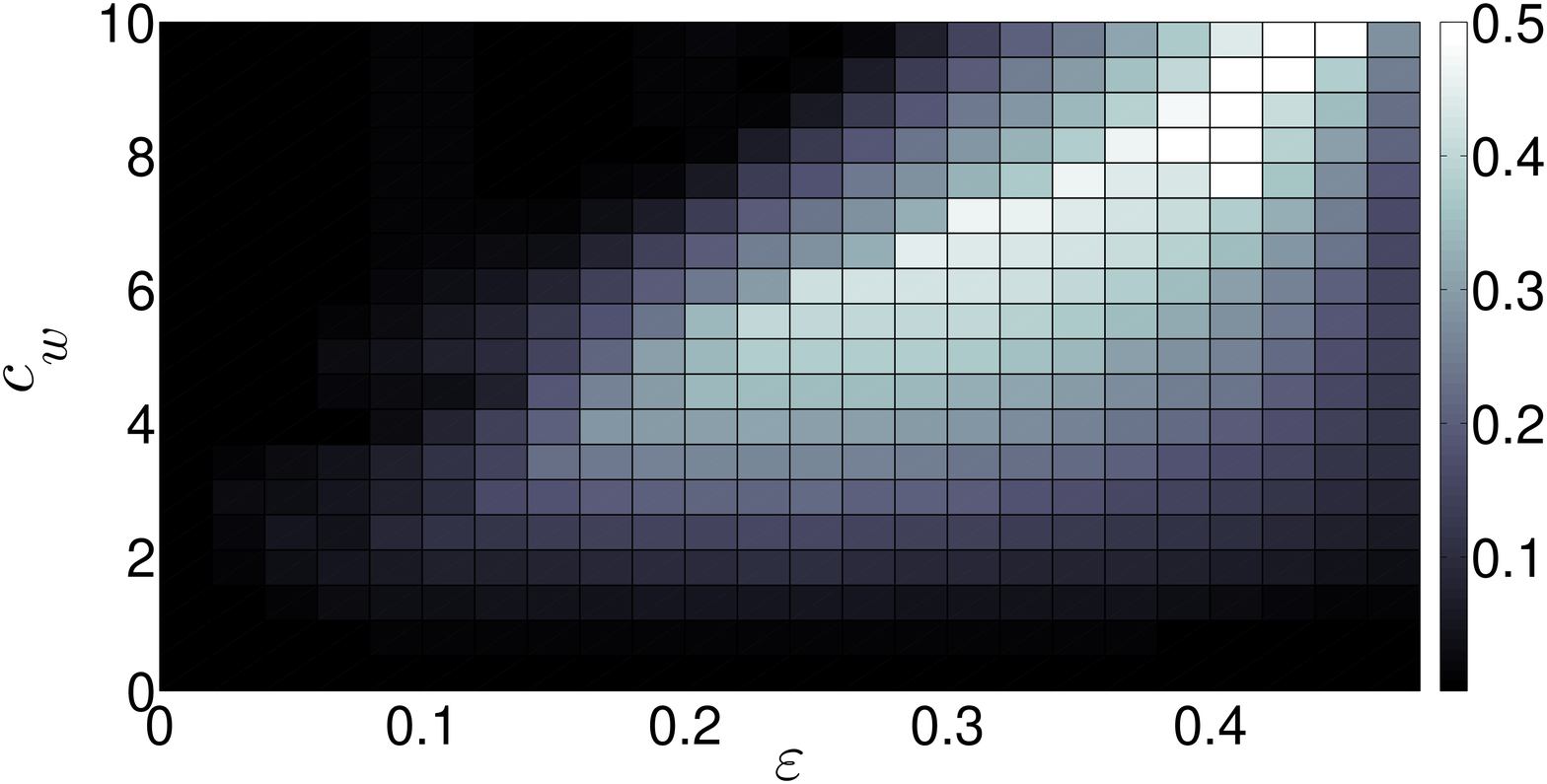}
     \includegraphics[width=0.55\textwidth]{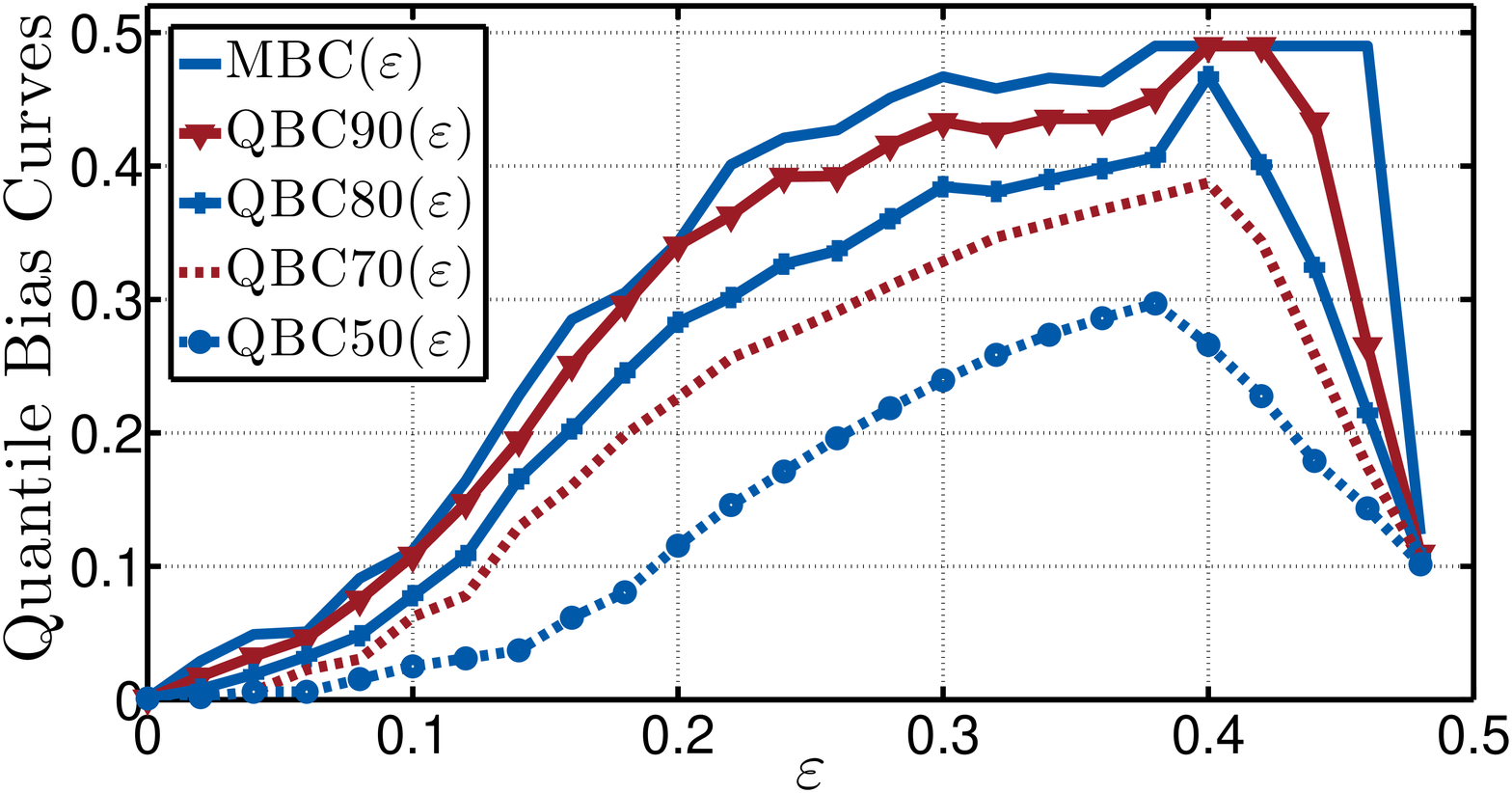} 
     \caption{(left) The maximum bias and quantile bias curves of the BIP $\tau$-estimator for the AR(1) with AOs. The top graph shows the maximum bias for a given pair of $(c_w,\varepsilon)$. The bottom plot represents the QBC obtained assuming the worst possible $c_w$ for a fixed $\varepsilon$.}
     \label{fig:dependent-bip-tau-qbc-ar1}
 \end{figure}
\else
\begin{figure}[htp]
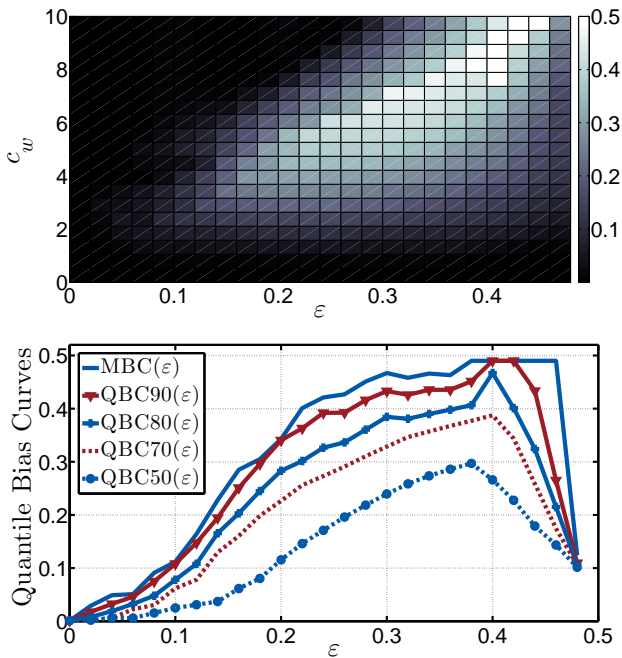

   \centering
     \includegraphics[width=0.5\textwidth]{fig4.eps}
     \includegraphics[width=0.5\textwidth]{fig4b.eps} 
      \vspace{-25 pt}
     \caption{(left) The maximum bias and quantile bias curves of the BIP $\tau$-estimator for the AR(1) with AOs. The top graph shows the maximum bias for a given pair of $(c_w,\varepsilon)$. The bottom plot represents the QBC obtained assuming the worst possible $c_w$ for a fixed $\varepsilon$.}
     \label{fig:dependent-bip-tau-qbc-ar1}
 \end{figure}

\fi
\subsection{Comparison to existing robust methods}
Our proposed estimator is compared numerically to the following methods.

\paragraph*{3$\sigma$ cleaned ML-estimator (ML $3\sigma$)}
This estimator is a simple diagnostic robust method that is frequently used among engineering practitioners \cite{zoubir-2012}. It applies an ML-estimator after a {\it 3$\sigma$ rejection}, i.e., observations beyond three standard deviations are flagged as outliers. 
 In this implementation, the median and the normalized median absolute deviations estimators of location and scale and the ML ARMA-estimator by Jones \cite{jones-1980} are used.

\paragraph*{BIP MM-estimator}
The BIP MM-estimator is a sophisticated robust estimator that has been proposed by Muler {\it et al.} \cite{muler-2009} who introduced the BIP model. MM-estimation consists of computing in the first step a highly robust estimate of the error scale, and in the second step, using this scale estimate to compute an efficient M-estimate. Its performance strongly depends on the starting point. 

\paragraph*{Filtered $\tau$-estimator (Filt $\tau$)}
\label{para:filtered-est}
An alternative approach to prevent the propagation of outliers is to combine robust estimators with {\it approximate conditional mean (ACM) type filters} (see \cite{masreliez-1975, martin-1982, spangl-2007,maronna-2006}). 
As a benchmark comparison, the {\it filtered $\tau$-estimator} is considered. This estimator finds the estimates $\hat{\boldsymbol{\beta}}$ such that the $\tau$-scale-estimate of the filtered innovations sequence is minimized. See \cite{maronna-2006} for a detailed discussion of this estimator. 


{\it Implementation} The implementation for the benchmark comparison in the case of the ML and the 3$\sigma$ cleaned ML is straightforward. For the BIP MM \cite{muler-2009} and the Filt $\tau$ \cite{maronna-2006}, no code is publicly available and the performance strongly depends on the starting point, which cannot be found by a grid search for the model orders considered. To provide a fair comparison, these methods are initialized with the same starting point as the BIP $\tau$. To verify the correctness of our implementations of these methods, we reproduced the experiments conducted in \cite{muler-2009} and obtained similar results for the BIP MM. For the Filt $\tau$, performance in the case of ARMA models could not be obtained as reported in \cite{muler-2009, maronna-2006}.  For this case, only the Filt $\tau$ results for the AR models are displayed, where the correctness of the implementation could be verified by comparing results to those published in \cite{muler-2009, maronna-2006}.
\vspace{-10 pt}
\subsection{Monte Carlo study on bias and standard deviation}
Next, numerical experiments to assess the average performance in terms of the bias and standard deviation for some ARMA models with $4\leq p+q\leq 8$ are conducted. In all cases, results represent averages over 1000 Monte Carlo runs. Presenting results for such ranges of $p,q$ is unusual in robust ARMA parameter estimation, which usually considers ARMA models of lower orders \cite{martin-1986,muler-2009, maronna-2006,chakhchoukh-2009c, chakhchoukh-2010t}. For our proposed estimator, $\rho_1$ and $\rho_2$ are chosen as in \eqref{eq:tau-rho} with two choices of $c_1$, as listed in Tables~\ref{table:example_ar4}-\ref{table:example_arma44} and $\Delta_{\zeta^0}=0.05$. The forward-backward algorithm and the initial starting point for the ARMA are abbreviated by fb and init, respectively. To be able to compute the Filt $\tau$ and BIP MM for such models, both methods are initialized with a starting point that was determined by our proposed robust starting point algorithm.

In our experiments, both patchy and independent replacement and AOs of different types are considered. Best average performance, i.e., best $\mu_{\hat{\beta}}$ is highlighted in bold font. Small standard deviations are only a useful measure of performance if the estimator does not break down, since breakdown can mean that all estimates take a similar (false) value. For this reason, $\mu_{\hat{\beta}}$ and $\sigma_{\hat{\beta}}$ are displayed, instead of mean-squared errors, in Tables~\ref{table:example_ar4}-\ref{table:example_arma44}.

{\noindent \it Example AR(4): $\boldsymbol{\phi}=(-2.7607,3.8106,-2.6535,0.9238)$, $\sigma=1$, $\mu=0$, $n=75$}\\
 This model was investigated for the clean data case in \cite{mcquarrie-1998}. $\mathrm{AO}_1$ refers to a single AO ($\varepsilon=0.0133$), where $w_t\sim\mathcal{N}(0,\sigma_w^2)$ with $\sigma_{w}=5\sigma_a$. $\mathrm{RO}_1$ refers to a single replacement outlier ($\varepsilon=0.0133$), where $w_t\sim\mathcal{N}(0,\sigma_w)$ with $\sigma_{w}=5\sigma_a$. $\mathrm{PAO}_{20}$ refers to large positive patchy AOs (patch length = 20, i.e., $\varepsilon=0.2667$), where $w_t\sim|\mathcal{N}(0,\sigma_w^2)|$ with $\sigma_{w}=5\sigma_x$. $\mathrm{PRO}_{20}$ on the other hand considers positive patchy replacement outliers (patch length = 20, i.e., $\varepsilon=0.2667$) whose standard deviation is identical to the uncorrupted process, where $w_t\sim|\mathcal{N}(0,\sigma_w^2)|$ with $\sigma_{w}=\sigma_x$. This is aparticularly challenging case.

Table~\ref{table:example_ar4} summarizes the results. As could be expected, the ML and ML 3$\sigma$ only perform well in the clean data case, i.e., $y_t=x_t$. The Filt $\tau$-estimator performs reasonably well, but is outperformed by all BIP estimators. The performance difference between the BIP $\tau$- and the BIP MM-estimators is not significant, which is reasonable, since they use the same starting point. Best performance depends on the type of outliers.

\if\paper\singlecol
{\noindent \it Example AR(7): $\boldsymbol{\phi}=(-3.5258, 6.9530, -9.3074,8.9473, -6.1572, 2.8428,-0.7059)$, $\sigma=1$, $\mu=0$, $n=50$}\\
\else
{\noindent \it Example AR(7): $\boldsymbol{\phi}=(-3.5258, 6.9530, -9.3074,8.9473,$ $-6.1572, 2.8428,-0.7059)$, $\sigma=1$, $\mu=0$, $n=50$}\\
\fi
The frequency response obtained with these parameters corresponds to that of a Hamming-window based linear-phase filter with normalized cutoff frequency at 0.5. $\mathrm{AO}_1$, $\mathrm{AO}_2$ and $\mathrm{AO}_3$ refer to 1, 2, and 3 isolated AOs whose distribution is $w_t\sim\mathcal{N}(0,\sigma_w^2)$ with $\sigma_{w}=\sigma_x$. 

Table~\ref{table:example_ar7} summarizes the results. As for the previous experiment, the MLE performs best for the clean data case and the BIP model based estimators provide best performance in the presence of outliers. In this experiment, the BIP $\tau$ consistently outperforms its robust competitors for all considered scenarios.

%
%

{\noindent \it ARMA(4,4): $\boldsymbol{\phi}=(0.100,1.6600,0.0930,0.8649)$, $\boldsymbol{\theta}=(0.0226,0.8175,0.0595,0.0764)$, $\sigma=1$, $\mu=0$, $n=1000$}\\
This model was investigated for the clean data case in \cite{moses-1987}. The data is contaminated by independent AOs, with $w_t\sim\mathcal{N}(0,\sigma_w^2)$ where $\sigma_w=10$. 

Table~\ref{table:example_arma44} summarizes the results. As in the previous experiments, the BIP model based estimators exhibit a good resistance against outliers (in this case up to 40 percent) and also perform well for the clean data case. Table~\ref{table:example_arma44} also displays the robust starting point $\hat{\boldsymbol{\beta}}_0$, for which an AR($8$) approximation was used. In this example, because the outliers are easily detected by the 3$\sigma$ rule, the performance of the 3$\sigma$ ML is surprisingly good up to $\varepsilon=0.25$.
\vspace{-10 pt}
\subsection{Choice of AR order in the robust starting point algorithm}
\label{subsec:ar_order_approx}
Fig.~\ref{fig:choice_of_p_for_ARMA44} plots the Monte Carlo averaged mean absolute error of the ARMA(4,4) parameter estimates for the above example as a function of the order of the AR approximation that is used to find the starting point. For the clean data-case, the choice of the order is not critical, since $y_t^* \approx y_t$, i.e., not much outlier cleaning is performed for any of the AR models. In the case of additive outliers the order should be chosen large enough so that the cleaned values, i.e., the values for which $y_t^*\neq y_t$, approximately fit into an ARMA(4,4) model. In practice, numerical experiments suggest a value in the range of $p+q\leq p \leq 2(p+q)$ is sufficient to find a starting point.\\ 
\if\paper\singlecol
\begin{figure}[htbp]
   \centering
     \includegraphics[width=0.55\textwidth]{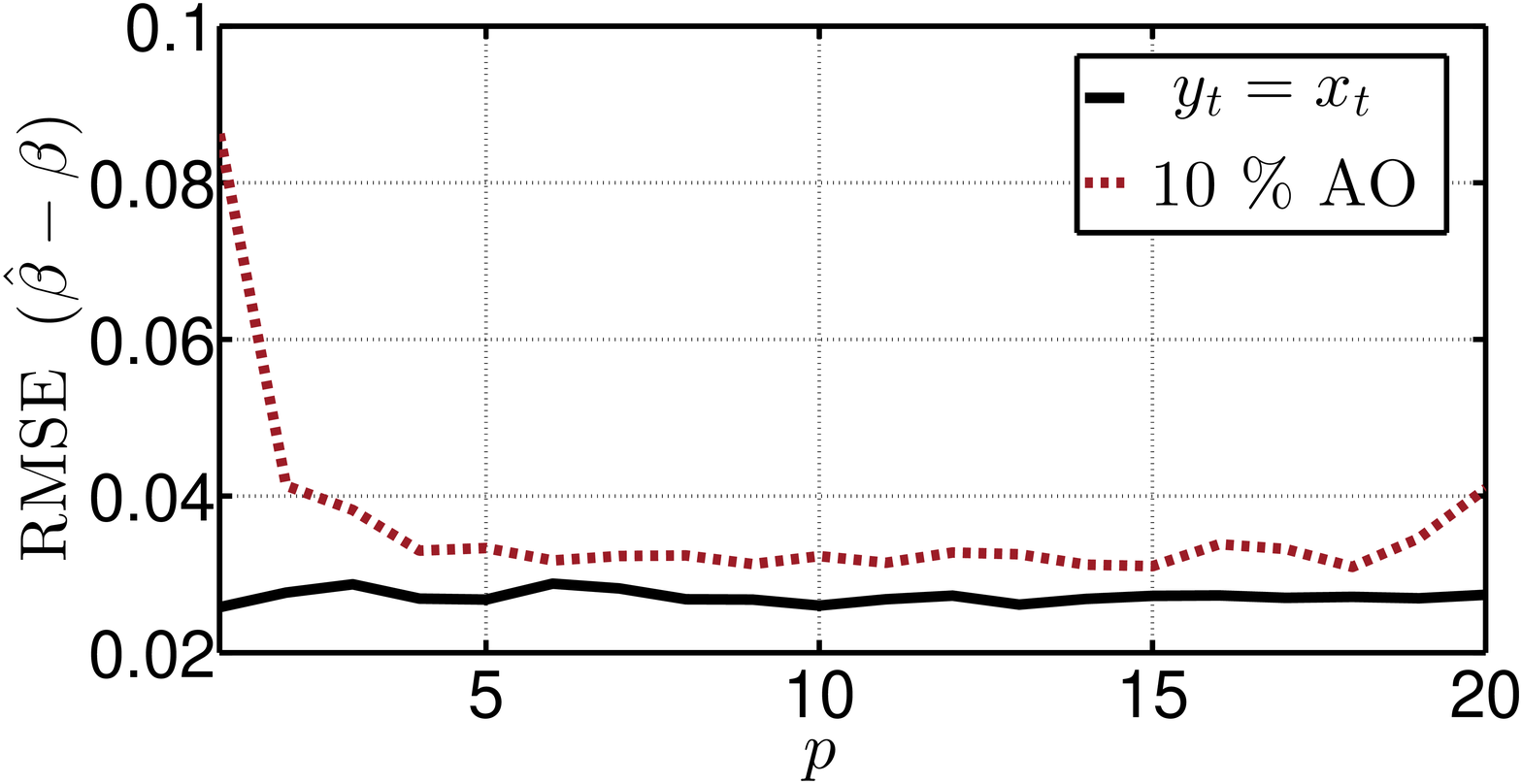}
\vspace{-25 pt}
     \caption{Evaluation of the effect of the AR order on the final estimates.}
     \label{fig:choice_of_p_for_ARMA44}
 \end{figure}
\else
\begin{figure}[htbp]
   \centering
     \includegraphics[width=0.5\textwidth]{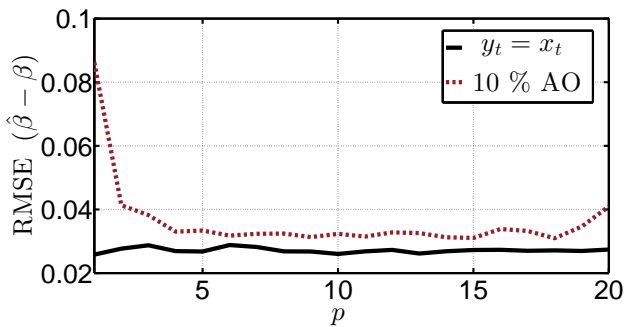}
\vspace{-25 pt}
     \caption{Evaluation of the effect of the AR order on the final estimates.}
     \label{fig:choice_of_p_for_ARMA44}
 \end{figure}
\fi

\subsection{Computational complexity of the algorithm}
\label{subsec:complexity}
Allthough a theoretical complexity analysis of the proposed algorithm cannot be derived, information deduced from Monte-Carlo averaged runtimes, is useful. Firstly, the main computation time (on average 82.925 \% for the ARMA(4,4) example), is required to find a robust starting point, since Marquard algorithms can solve nonlinear LS problems very efficiently. Therefore, the focus of complexity analysis is on the AR parameter estimation. Results are displayed for Algorithm 1; runtimes for Algorithm 2 are approximately twice as long. Secondly,  the computational complexity of robust methods far exceeds that of non-robust methods. Thirdly, the runtimes of the algorithms strongly depend on the available processing power\footnote{The presented average runtimes are based on an Intel Core i5 CPU 760, 2.80 GHz x 4, where no parallel multicore processing has been performed.}, and, accordingly, the relative differences are of more important interest than the absolute values.

Fig.~\ref{delta_zeta_0} displays the reduction of computation time that is achieved by first evaluating (\ref{eq:tau-arma-inno-scale}) and (\ref{eq:tau-bip-arma-inno-scale}) on a coarse grid and then interpolating the curves onto a grid of $\Delta_{\zeta}=0.001$ by using a least-squares polynomial fit of order four compared to evaluating  (\ref{eq:tau-arma-inno-scale}) and (\ref{eq:tau-bip-arma-inno-scale}) directly on $\Delta_{\zeta}=0.001$. 
\if\paper\singlecol
\begin{figure}[htbp]
\centering
\includegraphics[width=0.55\textwidth]{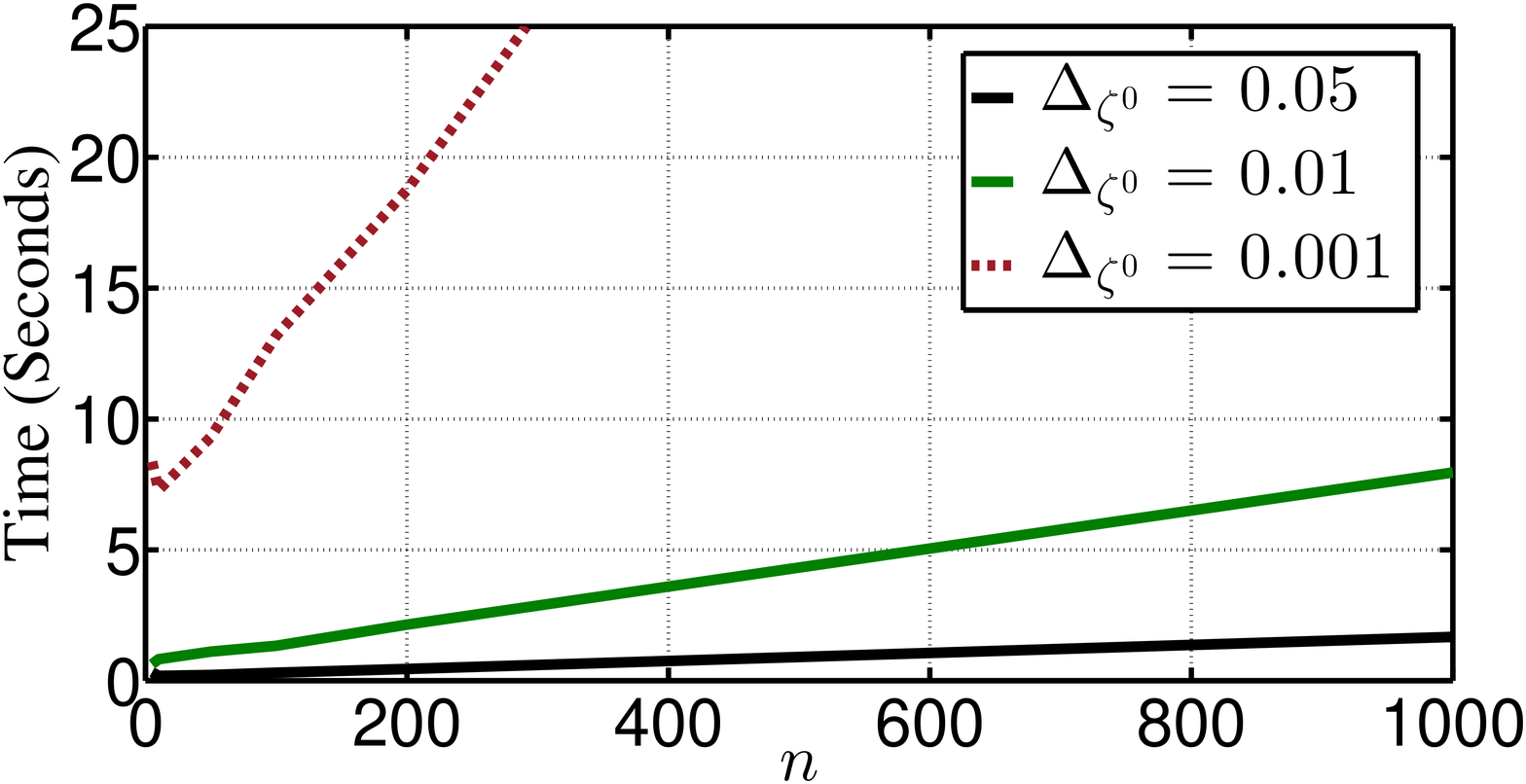}
\vspace{-10 pt}
\caption{Computation times for $p=1$ for different step sizes $\Delta_{\zeta^0}$ of the grid on which the $\tau$-scale is evaluated.}
\label{delta_zeta_0}
\end{figure}
\else
\begin{figure}[htbp]
\centering
\includegraphics[width=0.5\textwidth]{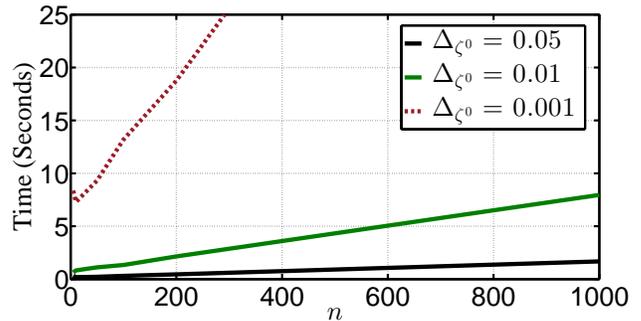}
\vspace{-10 pt}
\caption{Computation times for $p=1$ for different step sizes $\Delta_{\zeta^0}$ of the grid on which the $\tau$-scale is evaluated.}
\label{delta_zeta_0}
\end{figure}
\fi

Fig.~\ref{fig:computation_time_n_p} plots the computation times for different AR model orders $p$ as a function of the sample size $n$. The increase is a linear function of $n$. Further numerical experiments, which are not reported here due to space limitations, show that the complexity for a fixed sample size is also linearly related to $p$.
\if\paper\singlecol
\begin{figure}[htbp]
\centering
\includegraphics[width=0.55\textwidth]{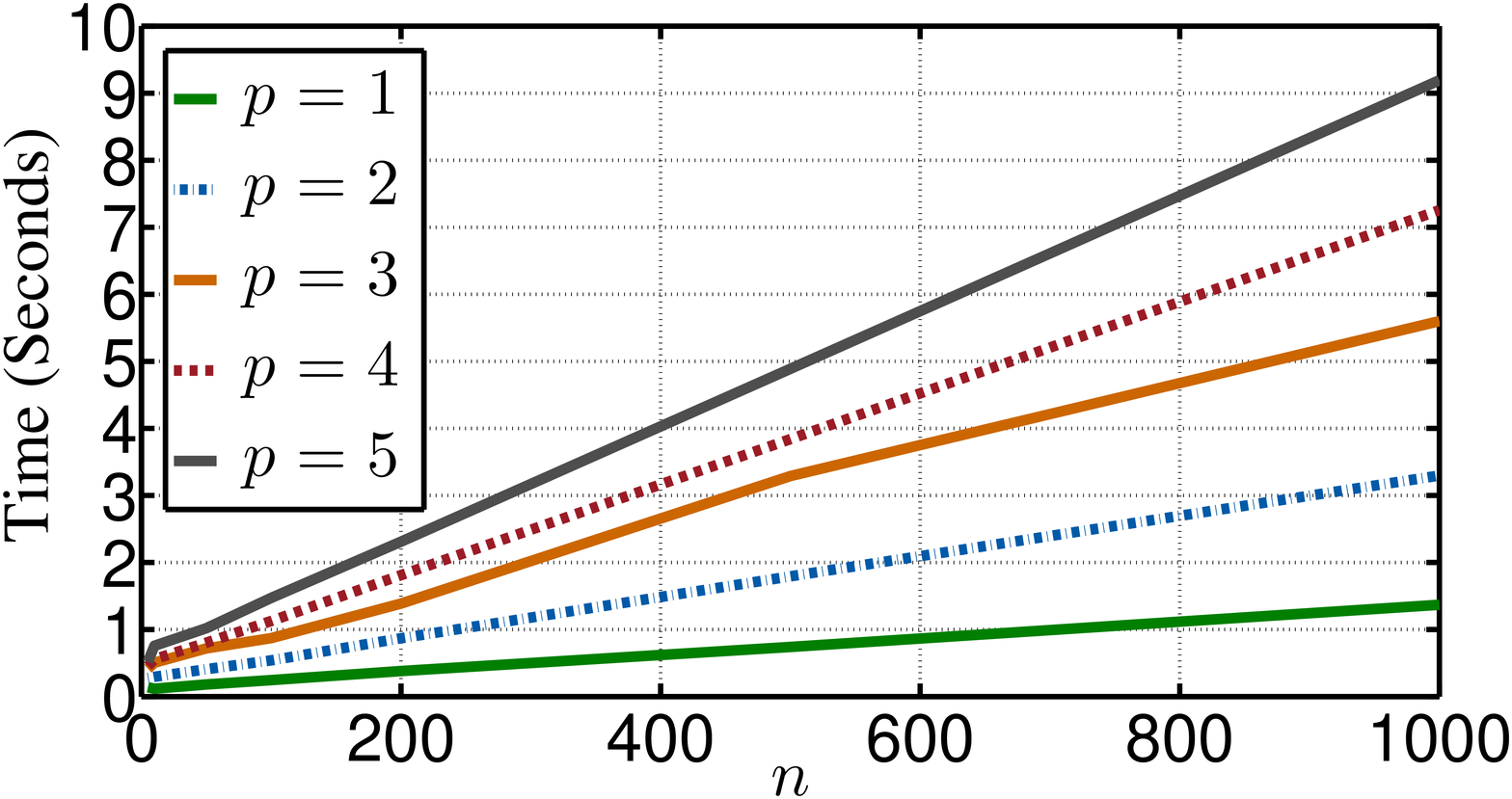}
\vspace{-10 pt}
\caption{Computation times for different AR model orders $p$ as a function of the sample size $n$. $\Delta_{\zeta^0}=0.05$}
\label{fig:computation_time_n_p}
\end{figure}
\else
\begin{figure}[htbp]
\centering
\includegraphics[width=0.5\textwidth]{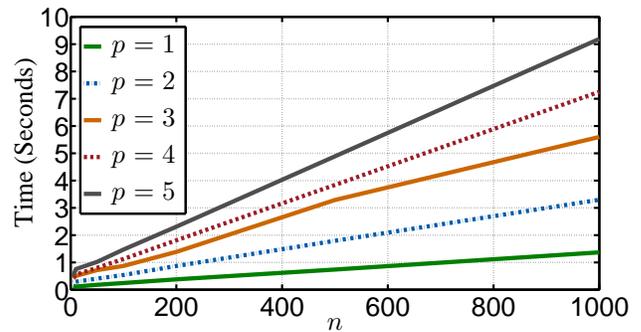}
\vspace{-10 pt}
\caption{Computation times for different AR model orders $p$ as a function of the sample size $n$. $\Delta_{\zeta^0}=0.05$}
\label{fig:computation_time_n_p}
\end{figure}
\fi

\vspace{-5 pt}
\section{Real-Data Example}
\label{sec:real-example}
Finally, the real-data applicability of our proposed estimator is illustrated by considering the practical application of cleaning the R-R interval plots from errors that are introduced by imperfections of an R-peak detection algorithm. The ECG data that is shown, is part of a larger dataset that was recorded at Technische Universit\"at Darmstadt in cooperation with the Department of Psychology using the Biopac MP 150 System and the AcqKnowledge 4.2 Software (Biopac Systems, 2011). The data was sampled with a sampling frequency of 250 Hz. To extract the R-R intervals, the QRS detector by Pan and Tompkins \cite{pan-1985} that was implemented by Clifford \cite{clifford-2002} was applied. As can be seen from Fig.~\ref{fig:hrv_example} (top), most of the R-peaks of the ECG were correctly detected, however, because of some occasional misdetections and false alarms, the R-R interval series contains outliers. 

\if\paper\singlecol
\begin{figure}[htbp]
   \centering
     \includegraphics[width=0.6\textwidth]{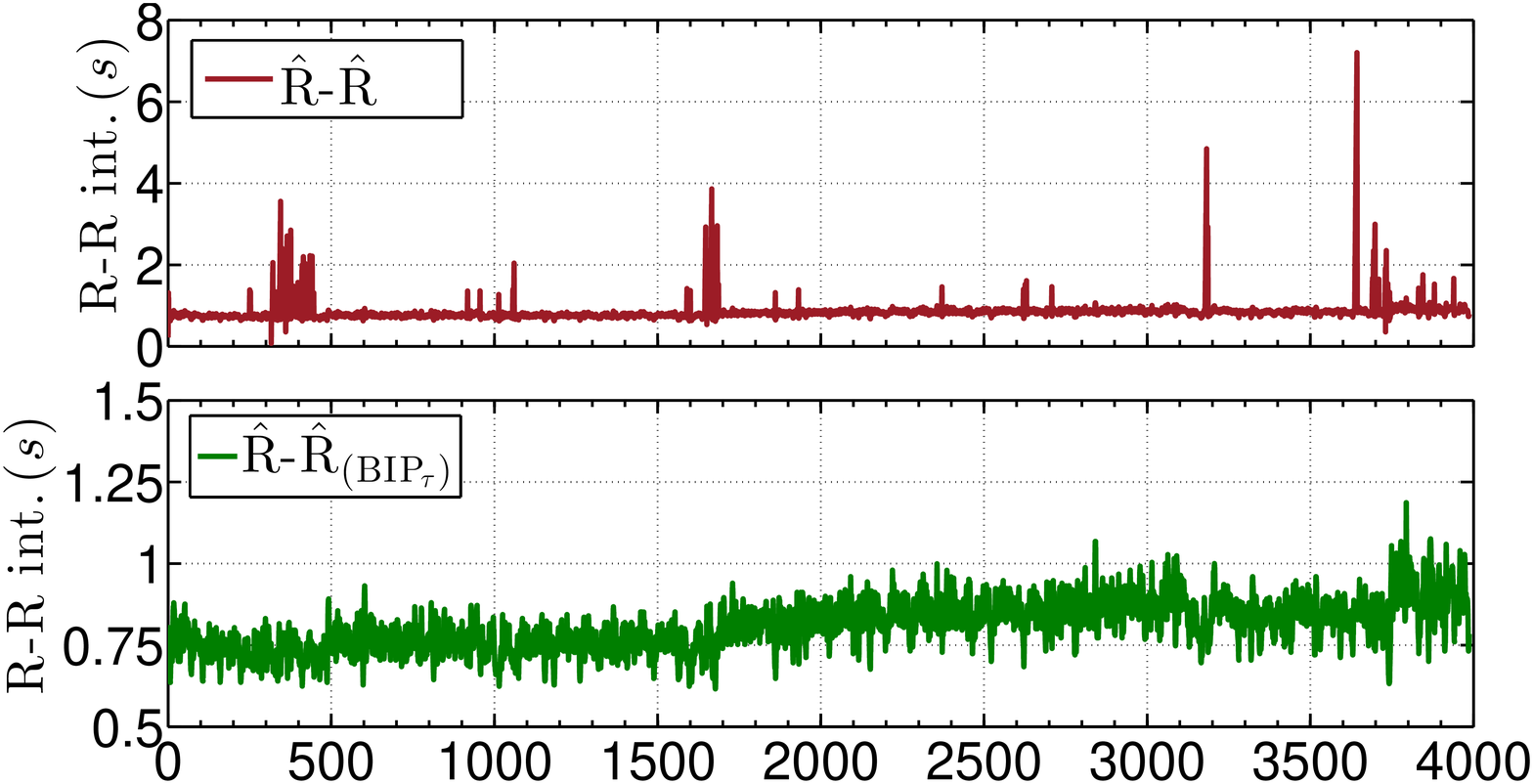}
     \includegraphics[width=0.6\textwidth]{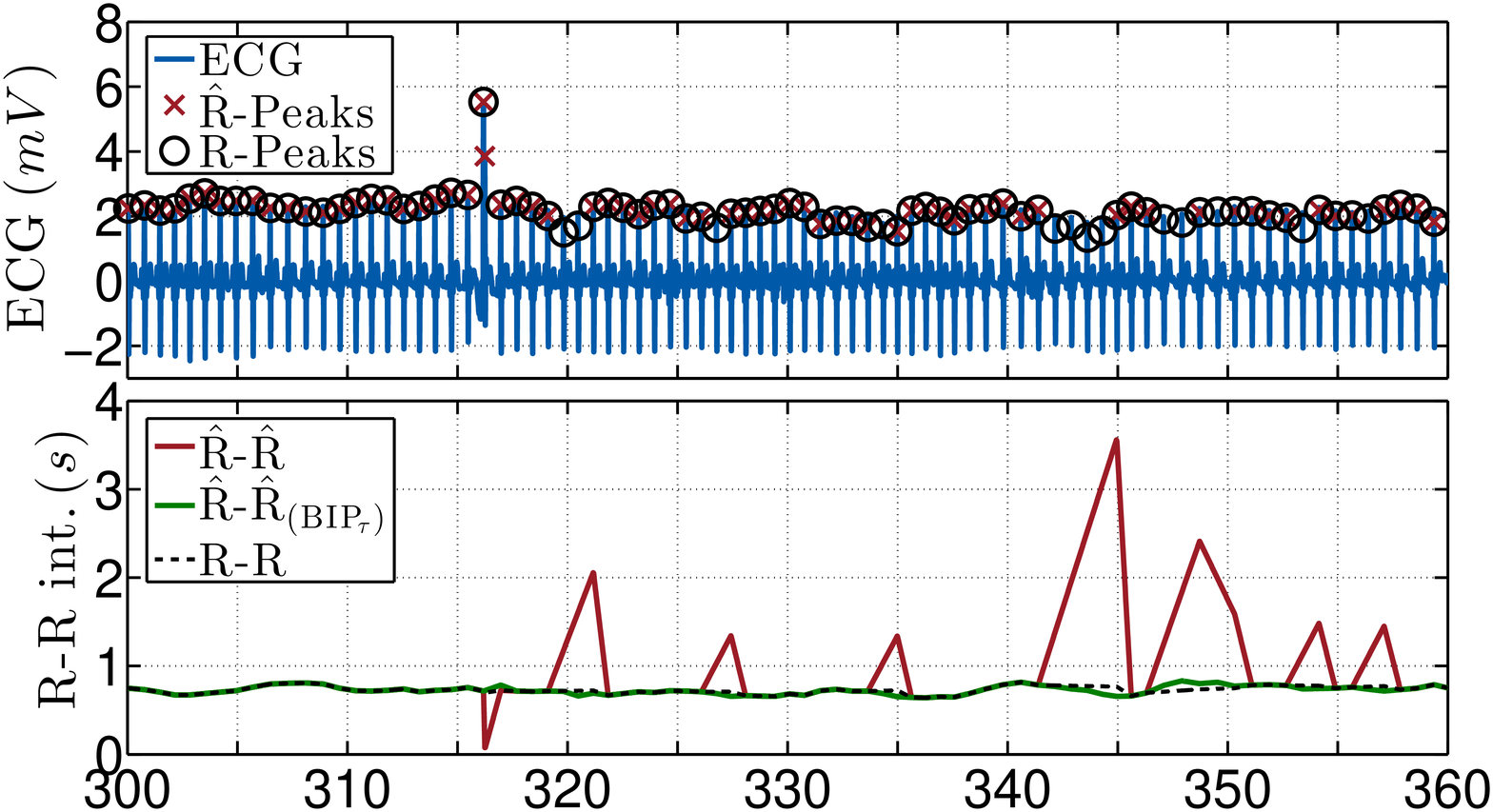} 
     \caption{An application of the proposed estimator for cleaning of R-R interval plots. From these plots HRV metrics are derived. The bottom two plots show a 60 second except containing ground truth R-peaks and cleaned R-R plots.}
     \label{fig:hrv_example}
 \end{figure}
\else
\begin{figure}[htbp]
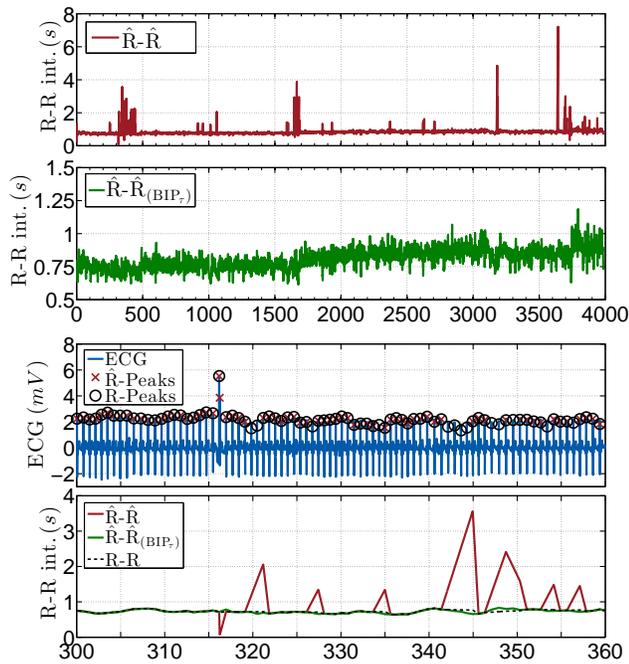

   \centering
     \includegraphics[width=0.5\textwidth]{fig8.eps}
     \includegraphics[width=0.5\textwidth]{fig8b.eps} \\ 
    \vspace{-5 pt}
     \caption{An application of the proposed estimator for cleaning of R-R interval plots. From these plots HRV metrics are derived. The bottom two plots show a 60 second except containing ground truth R-peaks and cleaned R-R plots.}
     \label{fig:hrv_example}
 \end{figure}
\fi

The proposed estimator was used to outlier clean the R-R interval series by applying \eqref{eq:bip-cleaning} and using an AR(5) model. The result of the outlier cleaning is depicted in  Fig.~\ref{fig:hrv_example} (2nd from top). To determine the correct model order, i.e., to estimate $p$, robust model order selection criteria \cite{muma.ssp-2014} were applied based on the final $\tau$-estimate of the innovations scale, i.e.,
\begin{equation}
 \mathrm{IC}(p) = \log(\hat{\sigma}_\tau^*(p)^2)+c_{\mathrm{pen}} p.
\end{equation}
The results of the robust model order selection are provided in Table \ref{table:mos_rr}. By choosing $c_{\mathrm{pen}}=2(p+1)/n$, $c_{\mathrm{pen}}=\log(n)p/n$ and $c_{\mathrm{pen}}=2\log(\log(n))p/n$ the criteria by Akaike, Schwarz and Hannan and Quinn, stated respectively in \cite{muma.ssp-2014, mcquarrie-1998} are obtained. The third plot of  Fig.~\ref{fig:hrv_example} details a particular outlier contaminated region of the R-R series, for which we have manually corrected the R-peak detection to obtain a ground truth reference (black circles). The bottom plot displays the outlier cleaned R-R interval series (green), the original one derived from the faulty R-detection (red) and the one obtained from the ground truth R-peak detection (black). By comparing the plots, it becomes clear that, firstly, only the outlying R-intervals are cleaned, and secondly, the correction is close to the ground truth value. The chosen example is typical of the results obtained for the entire data set. The full dataset and the 
Matlab 
code to reproduce Fig.~\ref{fig:hrv_example}, are available upon request.
\begin{table*}[ht]
\footnotesize
\centering
\vspace{-5 pt}
\caption{Robust BIP-$\tau$ based model order selection \cite{muma.ssp-2014} for the R-R interval series. The chosen order is highlighted with bold font.}
	\begin{tabular}{cccccccccccc}
	\toprule
	& 0 & 1 & 2 & 3 & 4  & 5 & 6  & 7 & 8  & 9 & 10\\
	\midrule
	AIC & -4.902 & -6.587 & -6.613 & -6.604  & -6.622 & {\bf -6.648}  & -6.631 & -6.658  & -6.575 & -6.612  & -6.621\\
	\midrule
	SIC & -4.902 & -6.586 & -6.611 & -6.601  & -6.617 & {\bf -6.641}  & -6.623 & -6.579  & -6.564 & -6.600   & -6.608\\
	\midrule
	HQC & -4.902 & -6.587 & -6.612 & -6.603  & -6.621 & {\bf -6.646}  & -6.628 & -6.658  & -6.571 & -6.608  & -6.616\\
	\bottomrule
\end{tabular}
\label{table:mos_rr}
\end{table*}

\vspace{-5 pt}
\section{Conclusion}
\label{sec:conclusion}
A new robust and statistically efficient estimator for ARMA models called the bounded influence propagation (BIP) $\tau$-estimator was proposed and analyzed. Strong consistency and asymptotic normality of the estimator for ARMA models that are driven by independently and identically distributed (iid) innovations with symmetric distributions were established. To analyze the infinitesimal effect of outliers on the estimator, the influence function was derived. The gross error sensitivity of the BIP $\tau$-estimator was found to be lower than that of existing robust estimators for an AR(1) with additive outliers. Algorithms were provided to compute the estimates. Numerical experiments were conducted to compare the finite sample performance of the proposed estimator to existing robust methodologies for different types of outliers both in terms of average and of worst-case performance, as measured by the maximum bias curve. A real-data example of outlier cleaning for R-R interval plots derived from 
electrocardiographic (ECG) data showed the practical applicability of the proposed estimator. The proposed estimator is also useful in many other real-world problems, e.g. speech processing, state estimation or econometry, which can be modeled as an ARMA that is disturbed by outliers or impulsive noise. Extensions to the Seasonal Integrated ARMA (SARIMA) or Periodic ARMA (PARMA) \cite{bondon-parma-2015} as well as vectorial AR (VAR) will be investigated in future.

\vspace{-5 pt}
\section*{Acknowledgments}
We thank the anonymous reviewers and Dr. Roy Howard for their careful reading of our manuscript and their many insightful comments and suggestions. This work was supported by the project HANDiCAMS which acknowledges the financial support of the Future and Emerging Technologies (FET) programme within the Seventh Framework Programme for Research of the European Commission, under FET-Open grant number: 323944.

\vspace{-5 pt}
\appendix[Results of the numerical experiments]
\begin{table*}[htbp]
\centering
\footnotesize
\caption{Monte Carlo experiment for the parameter estimation of an AR(4) with $\boldsymbol{\phi}=(-2.7607,3.8106,-2.6535,0.9238)$, $\sigma=1$, $\mu=0$, $n=75$. $c_{1,\mathrm{rob}}=0.8100$ and $c_{1,\mathrm{eff}}=0.4050$ (corresponding to 95 \% efficiency at the Gaussian ARMA model). Best performance in terms of bias for each parameter is highlighted with bold font.}
	\begin{tabular}{llcccccccccc}
	\toprule
 	\multicolumn{2}{c}{ }  & \multicolumn{2}{c}{$y_t=x_t$ } & \multicolumn{2}{c}{$\mathrm{AO}_1$ } & \multicolumn{2}{c}{$\mathrm{RO}_1$ } & \multicolumn{2}{c}{$\mathrm{PAO}_{20}$  }& \multicolumn{2}{c}{$\mathrm{PRO}_{20}$ }\\
	Parameter &Methods & $\mu_{\hat{\beta}}$ & $\sigma_{\hat{\beta}}$ & $\mu_{\hat{\beta}}$ & $\sigma_{\hat{\beta}}$  & $\mu_{\hat{\beta}}$ & $\sigma_{\hat{\beta}}$  & $\mu_{\hat{\beta}}$ & $\sigma_{\hat{\beta}}$ & $\mu_{\hat{\beta}}$ & $\sigma_{\hat{\beta}}$\\
	\midrule
	&ML & -2.7272 & 0.0688 & -2.2174  & 0.4292 & -1.2482  & 0.5832 & -0.5266  & 0.1746 & -0.6473  &  0.1665\\
	&ML 3$\sigma$& -2.5130 & 0.5603 & -2.0327  & 0.5837 & -1.2186  & 0.5705 & -0.8544 & 0.7296 & -0.8150  & 0.2322\\
	&MRE& -0.5599 & 7.7694 & -1.8573  & 39.5711 & -1.1769  & 23.5192 & -0.7473  & 4.2561 & -0.9731 & 25.7938\\
	$\phi_1=-2.7607$&BIP MM & -2.7708 & 0.3543 & {\bf -2.7554}  & 0.3693 & -2.7376  & 0.3787 & -2.5936  & 0.5879& -2.5145  & 0.6385\\
	&Filt $\tau$& -2.4317 & 0.5991 & -2.2119  & 0.6504 & -2.3171 & 0.6310 & -1.6414  & 0.4637& -1.3558  & 0.4685\\
	&BIP $\tau$ $c_{1,\mathrm{rob}}$ & -2.8001 & 0.3791 & -2.7729 & 0.3868 &  {\bf -2.7519}  & 0.4146 &  {\bf -2.6086} & 0.6220&  {\bf -2.5230}  & 0.6720\\
	&BIP $\tau$ $c_{1,\mathrm{eff}}$ & {\bf -2.7622} & 0.2625 & -2.7225  & 0.2758 & -2.7104  & 0.3016& -2.4561  & 0.7296& -2.4326  & 0.7028\\
	&BIP $\tau$ $c_{1,\mathrm{eff,fb}}$& -2.1619 & 0.7679 & -2.0896  & 0.7534 & -1.9569  & 0.7430 & -1.8346 & 0.7139 & -1.8537  & 0.6692 \\
	\midrule
	&ML &  {\bf 3.7188} & 0.1628 & 2.5446  & 0.9661 & 0.8504  &  0.9615 & -0.1449   & 0.2139 & -0.0556  & 0.2135\\
	&ML 3$\sigma$& 3.3230 & 1.0152 & 2.0327  & 1.1239 & 0.8198  & 0.9068 & 0.4006  & 0.7454& 0.1723  & 0.3748\\
	&MRE& -0.0155 & 10.2315& 0.6752  & 23.0591 & 1.9330  & 51.8618 & 0.0027  & 8.8627& -0.8450  & 38.5007\\
	$\phi_2=3.8106$ &BIP MM & 3.6821 & 0.5565 & 3.6407  & 0.5737 & 3.6323  & 0.6037 & 3.3832  & 1.0585& 3.2687  & 1.1083\\
	&Filt $\tau$& 3.0866 & 1.1622 & 2.6242  & 1.2627 & 2.8394  & 1.2594 & 1.3569  & 0.9199& 1.0812  & 0.7998\\
	&BIP $\tau$ $c_{1,\mathrm{rob}}$ & 3.7008 & 0.5765 &  {\bf 3.6508}  & 0.5857 &  {\bf 3.6408} & 0.6232 &  {\bf 3.3934}  & 1.0608&  {\bf 3.2737} & 1.1122\\
	&BIP $\tau$ $c_{1,\mathrm{eff}}$ & 3.6940 & 0.3737 & 3.6187  & 0.3974 & 3.6186  & 0.4516& 3.1170  & 1.2694& 3.0599 & 1.1247\\
	&BIP $\tau$ $c_{1,\mathrm{eff,fb}}$& 2.5227 & 1.5461 & 2.4165  & 1.4910 & 2.1916  & 1.5078 & 1.7183 & 1.5822 & 1.8517  & 1.5030 \\	
	\midrule
	&ML &  {\bf -2.5587} & 0.1659 & -1.4027 & 0.9405 &   0.0085  & 0.7971 & -0.0251  & 0.2198 & 0.0920  & 0.2243\\
	&ML 3$\sigma$& -2.2157 & 0.8773 & -1.1477  & 1.0241 & 0.0179  & 0.7421 & -0.0512  & 0.6078& 0.1490  & 0.3408\\
	&MRE& 0.3785 & 10.2031 & 1.0557  & 23.1040 & -1.7660  & 51.8481 & 0.1724  & 8.9325 & 2.0991  & 38.4327\\
	$\phi_3=-2.6535$ &BIP MM & -2.4526 & 0.4686 & -2.4021  & 0.4761 & -2.4108  & 0.5139 &  {\bf -2.2155}  & 0.9490&  {\bf -2.1259}  & 0.9800\\
	&Filt $\tau$& -1.9435 & 1.1244 & -1.5171  & 1.2057 & -1.7128  & 1.2171 & -0.3875  & 0.9184& -0.2552  & 0.7627\\
	&BIP $\tau$ $c_{1,\mathrm{rob}}$ & -2.4317 & 0.4590 & -2.3896  & 0.4689 & -2.3965  & 0.5056 & -2.2099 & 0.9373& -2.1216  & 0.9725\\
	&BIP $\tau$ $c_{1,\mathrm{eff}}$ & -2.4726 & 0.2982 &  {\bf -2.4060}  & 0.3228 &  {\bf -2.4196}  & 0.3752 & -1.9697  & 1.1368& -1.9062 & 1.1457\\
	&BIP $\tau$ $c_{1,\mathrm{eff,fb}}$& -1.4176 & 1.5031 & -1.3325  & 1.4436 & -1.3322  & 1.4763 & -0.6067 & 1.6190 & -0.7734  & 1.5309 \\
	\midrule
	&ML &  {\bf 0.8804} &  0.0759 & 0.4233  & 0.3648 & 0.0999  & 0.2575 & -0.0244  & 0.1625 & -0.0159  & 0.1560\\
	&ML 3$\sigma$& 0.7843 & 0.2596 & 0.3675 & 0.3648 & 0.1084  & 0.2689 & 0.1775 & 0.3107& 0.0232 & 0.2102\\
	&MRE& 0.3152 & 7.7540 & -0.9727  & 39.5902 & 1.5158  & 23.4962 & 0.3240  & 4.3423 & -1.1864  & 25.7253\\
	$\phi_4=0.9238$ &BIP MM & 0.7683 & 0.1970 & 0.7465  & 0.2000 & 0.7644  & 0.2091 &  {\bf 0.7169}  & 0.3658&  {\bf 0.7099}  & 0.3743\\
	&Filt $\tau$& 0.6457 & 0.4695 & 0.4970  & 0.4964 & 0.5592  & 0.5115 & -0.0175  & 0.4697 & -0.004  & 0.4027\\
	&BIP $\tau$ $c_{1,\mathrm{rob}}$ & 0.7445 & 0.1892 & 0.7337  & 0.1951 & 0.7506  & 0.1999 & 0.6981  & 0.3606& 0.6906  & 0.3594\\
	&BIP $\tau$ $c_{1,\mathrm{eff}}$ & 0.7967 & 0.1516 &  {\bf 0.7773}  & 0.1630 &  {\bf 0.7932}  & 0.1687 & 0.6255  & 0.4343& 0.6001  & 0.4464\\
	&BIP $\tau$ $c_{1,\mathrm{eff,fb}}$& 0.4166 & 0.6525 & 0.4050  & 0.6285 & 0.3469  & 0.6480 & 0.0618 & 0.7886 & 0.1608  & 0.7461 \\
	\bottomrule
	\end{tabular}
\label{table:example_ar4}
\end{table*}
\begin{table*}[htpb]
\centering
\footnotesize
\caption{Monte Carlo experiment for the parameter estimation of an AR(7) with $\boldsymbol{\phi}=(-3.5258, 6.9530, -9.3074,8.9473, -6.1572, 2.8428,-0.7059)$, $\sigma=1$, $\mu=0$, $n=50$. $c_{1,\mathrm{rob}}=0.8100$ and $c_{1,\mathrm{eff}}=0.4050$ (corresponding to 95~\% efficiency for the Gaussian ARMA model). Best performance in terms of bias for each parameter is highlighted with bold font.}
	\begin{tabular}{llccccccccc}
	\toprule
 	\multicolumn{2}{c}{ }  & \multicolumn{2}{c}{$y_t=x_t$ ($\varepsilon=0$)} & \multicolumn{2}{c}{$\mathrm{AO}_1$ ($\varepsilon=0.02$)} & \multicolumn{2}{c}{$\mathrm{AO}_2$ ($\varepsilon=0.04$)} & \multicolumn{2}{c}{$\mathrm{AO}_{3}$ ($\varepsilon=0.06$)}\\	
	Parameter &Methods & $\mu_{\hat{\beta}}$ & $\sigma_{\hat{\beta}}$ & $\mu_{\hat{\beta}}$ & $\sigma_{\hat{\beta}}$  & $\mu_{\hat{\beta}}$ & $\sigma_{\hat{\beta}}$  & $\mu_{\hat{\beta}}$ & $\sigma_{\hat{\beta}}$\\
	\midrule
	&ML & {\bf -3.4353} & 0.1407 & -1.3891  & 0.0816 & -1.3292  & 0.0938 & -1.0565  & 0.0902 \\
	&ML 3$\sigma$& -2.7113 & 1.1828 & -1.0050  & 0.4860 & -1.1694  & 0.2662 & -0.9473 & 0.2023\\
	&BIP MM & -2.6524 & 0.9123 & -2.7141  & 0.8652& -2.5088 & 1.0249 & -2.3467  & 1.0882\\
	$\phi_1=-3.5258$ &Filt $\tau$& -2.2798 & 0.7397 & -1.8765  & 0.6726 & -1.3962 & 0.5125 & -1.2712  & 0.4129\\
	&BIP $\tau$ $c_{1,\mathrm{rob}}$ & -2.6628 & 0.8950 & -2.7067 & 0.8750 & -2.4900  & 1.0308 & -2.3384 & 1.1021\\
	&BIP $\tau$ $c_{1,\mathrm{eff}}$ & -3.0679 & 0.6986 & {\bf -2.9080} & 0.7939 & {\bf -2.6412}  & 0.9534 & {\bf -2.5042}  & 0.9971\\
	&BIP $\tau$ $c_{1,\mathrm{eff,fb}}$ & -3.1519 & 0.5412 & -2.7187 & 0.7235 & -1.9790  &0.7817 & 1.7116  & 0.7821 \\
	\midrule
	&ML & {\bf 6.6280} & 0.4396 & 0.9911 & 0.1631 & 1.1240 &  0.0961 & 0.5221  & 0.1021 \\
	&ML 3$\sigma$& 4.9232 & 2.7819 & 0.6496  & 0.7536 & 0.9012  & 0.3927 & 0.4482  & 0.1939\\
	&BIP MM & 4.4563 & 2.3185 & 4.5704  & 2.2581 & 4.1243  & 2.6441 & 3.8317 & 2.6445\\
	$\phi_2=6.9530$ &Filt $\tau$& 3.3475 & 1.9313 & 2.4808  & 1.5172 & 1.3745 & 0.9546 & 1.1059  & 0.7301\\
	&BIP $\tau$ $c_{1,\mathrm{rob}}$ & 4.4540 & 2.3239 & 4.5626  & 2.2650 & 4.1149 & 2.6449 & 3.8091  &2.6659\\
	&BIP $\tau$ $c_{1,\mathrm{eff}}$ & 5.5738 & 1.9095 & {\bf 5.1319}  & 2.1251 & {\bf 4.4827}  & 2.5357 & {\bf 4.0770}  & 2.5960\\
	&BIP $\tau$ $c_{1,\mathrm{eff,fb}}$ & 5.7895 & 1.6229 & -4.4105 & 2.0701 & -2.7402  &1.9888 & 2.0826  & 1.9334 \\
	\midrule
	&ML & {\bf -8.6849} & 0.7835 & -0.0784 & 0.1940 &  -0.6310  & 0.1284 & 0.1217  & 0.1206\\
	&ML 3$\sigma$& -6.2725 & 3.9425 & -0.0908  & 0.6495 & -0.4192  & 0.4520 & 0.1149  & 0.2234\\
	&BIP MM &-5.2197 & 3.4983 & -5.3514  & 3.3969 & -4.8231 & 3.9715 & -4.4369  & 3.8166\\
	$\phi_3=-9.3074$ &Filt $\tau$& -3.3303 & 3.0250 & -2.1003  & 2.2103 & -0.7735  & 1.0865 & -0.4324  & 0.8485\\
	&BIP $\tau$ $c_{1,\mathrm{rob}}$ & -5.2245 & 3.4830 & -5.3466  & 3.4015 & -4.8279  & 3.9731 & -4.4499 & 3.8060\\
	&BIP $\tau$ $c_{1,\mathrm{eff}}$ & -6.9467 & 2.9714 & {\bf -6.2273} & 3.3143 & {\bf -5.3770}  & 3.8943 & {\bf -4.6962}  & 3.9031\\
	&BIP $\tau$ $c_{1,\mathrm{eff,fb}}$ & -7.2974 & 2.7028 & -4.9234 & 3.3977 & -2.7670  &2.9288 & -1.7982  & 2.7958 \\
	\midrule
	&ML & {\bf 8.1793} &  0.9335 & -0.3699  & 0.0907 & 0.5177  & 0.0930 & -0.0746  &0.0646\\
	&ML 3$\sigma$& 5.8583 & 3.8171 & -0.0873 & 0.4168 & 0.3654  & 0.3518 & -0.0467 & 0.1656\\
	&BIP MM & 4.5298 & 3.4874 & 4.5605  & 3.4649 & 4.1789  & 4.0830 & 3.8235  & 3.7295\\
	$\phi_4=8.9473$ &Filt $\tau$& 2.3825 & 3.2252 & 1.2127  & 2.2370 & 0.2622  & 0.8173 & 0.0128  & 0.7325 \\
	&BIP $\tau$ $c_{1,\mathrm{rob}}$ & 4.5028 & 3.5052 & 4.5408  & 3.4663 & 4.1722  & 4.0740 & 3.8067  & 3.7296\\
	&BIP $\tau$ $c_{1,\mathrm{eff}}$ & 6.2768 & 3.0783 & {\bf 5.4845}  & 3.4396 & {\bf 4.7711}  & 4.0299 & {\bf 3.9903}  & 3.9442\\
	&BIP $\tau$ $c_{1,\mathrm{eff,fb}}$ & 6.6373 & 2.9851 & 4.0069 & 3.6532 & -2.1874  &2.8919 & 1.2030  & 2.7626 \\
	\midrule
	&ML & {\bf -5.5059} & 0.7741 & 0.1616  & 0.2997 & -0.6736  & 0.1662 & -0.3720  & 0.1138\\
	&ML 3$\sigma$& -3.9388 & 2.6327 & -0.0157  & 0.4344 & -0.5616  & 0.2634 & -0.3264 & 0.2059\\
	&BIP MM & -2.8382 & 2.4965 & -2.7902  & 2.4891 & -2.6423  & 2.9502 & -2.4079  & 2.5409\\
	$\phi_5=-6.1572$ &Filt $\tau$& -1.1707& 2.4465 & -0.3738  & 1.6336 & -0.0792 & 0.6021 & 0.0131  & 0.6060\\
	&BIP $\tau$ $c_{1,\mathrm{rob}}$ & -2.8726 & 2.4500 & -2.8057 & 2.4849 & -2.6534  & 2.9478 & -2.4187 & 2.5386\\
	&BIP $\tau$ $c_{1,\mathrm{eff}}$ & -4.1037 & 2.1997 & {\bf -3.4956}  & 2.4699 & {\bf -3.1207}  & 2.9066 & {\bf -2.5316}  & 2.7367\\
	&BIP $\tau$ $c_{1,\mathrm{eff,fb}}$ & -4.3782 & 2.2606 & -2.4171 & 2.6873 & -1.3718  &2.0258 & -0.6731  & 1.9763 \\
	\midrule
	&ML & {\bf 2.4878} & 0.4318 & 0.2015  & 0.3534 & 0.6885 & 0.2012 & 0.5798  & 0.1165\\
	&ML 3$\sigma$& 1.7817 & 1.2299 & 0.0822  & 0.4210 & 0.5424 & 0.2838 & 0.4171  & 0.2795\\
	&BIP MM & 1.2617 & 1.1567 & 1.1540  & 1.1919 & 1.1253  & 1.4610 & 1.0264  & 1.1792\\
	$\phi_6=2.8428$ &Filt $\tau$& 0.3319& 1.2821 & -0.0237  & 0.8322 & 0.0661  & 0.4033 & 0.0819  & 0.4401\\
	&BIP $\tau$ $c_{1,\mathrm{rob}}$ & 1.2568 & 1.1616 & 1.1709  & 1.1992 & 1.1353 & 1.4474& 1.0199  & 1.1698\\
	&BIP $\tau$ $c_{1,\mathrm{eff}}$ & 1.8026 & 1.0441 & {\bf 1.5027}  & 1.1751 & {\bf 1.4048}  & 1.3931 & {\bf 1.1032} & 1.2651\\
	&BIP $\tau$ $c_{1,\mathrm{eff,fb}}$ & 1.9573 & 1.1562 & 0.9797 & 1.3511& 0.6580  &0.9747 & 0.3310  & 0.9772 \\
	\midrule
	&ML & {\bf -0.6006} & 0.1391 & -0.1786 & 0.2011 &  -0.3281  & 0.1671 & {\bf -0.2811}  & 0.1514 \\
	&ML 3$\sigma$& -0.4276 & 0.3322 & -0.0267  & 0.2700 & -0.2460  & 0.2013 & 0.1770 & 0.2146\\
	&BIP MM & -0.3060 & 0.3286 & -0.2717  & 0.3494 & -0.2709 & 0.4580 & -0.2297 & 0.3147\\
	$\phi_7=-0.7059$ &Filt $\tau$& -0.0315 & 0.3800 & 0.0746  & 0.8322 & -0.0290  & 0.2101 & -0.0712  & 0.2222\\
	&BIP $\tau$ $c_{1,\mathrm{rob}}$ & -0.3026 & 0.3220 & -0.2623  & 0.3372 & -0.2693  & 0.4156& -0.2315 &0.3227\\
	&BIP $\tau$ $c_{1,\mathrm{eff}}$ & -0.4275 & 0.2847 & {\bf -0.3498}  & 0.3150 & {\bf -0.3563} & 0.3952 & -0.2732  & 0.3408\\
	&BIP $\tau$ $c_{1,\mathrm{eff,fb}}$ & -0.4836 & 0.3390 & -0.2208 & 0.3966& -0.2053  &0.2939 & -0.1203  & 0.3004 \\
	\bottomrule
	\end{tabular}
\label{table:example_ar7}
\end{table*}
\begin{table*}[htpb]
\centering
\footnotesize
\caption{Monte Carlo experiment for the parameter estimation of an ARMA(4,4) with $\boldsymbol{\phi}=(0.100,1.6600,0.0930,0.8649)$, $\boldsymbol{\theta}=(0.0226,0.8175,0.0595,0.0764)$, $\sigma=1$, $\mu=0$, $n=1000$. $c_{1,\mathrm{rob}}=0.8100$ and $c_{1,\mathrm{eff}}=0.4050$ (corresponding to 95~\% efficiency for the Gaussian ARMA model). Best performance in terms of bias for each parameter is highlighted with bold font.}
	\begin{tabular}{llcccccccccc}
	\toprule
 	\multicolumn{2}{c}{ }  & \multicolumn{2}{c}{$y_t=x_t$} & \multicolumn{2}{c}{AO $\varepsilon=0.05$} & \multicolumn{2}{c}{AO $\varepsilon=0.10$} & \multicolumn{2}{c}{$\varepsilon=0.25$}& \multicolumn{2}{c}{AO $\varepsilon=0.40$}\\
	Parameter &Methods & $\mu_{\hat{\beta}}$ & $\sigma_{\hat{\beta}}$ & $\mu_{\hat{\beta}}$ & $\sigma_{\hat{\beta}}$  & $\mu_{\hat{\beta}}$ & $\sigma_{\hat{\beta}}$  & $\mu_{\hat{\beta}}$ & $\sigma_{\hat{\beta}}$ & $\mu_{\hat{\beta}}$ & $\sigma_{\hat{\beta}}$\\
	\midrule
	&ML & 0.0959 & 0.0187 & 0.0890  & 0.1426 & 0.0386  & 0.4413 & 0.0635  & 0.8285& -0.0527 & 0.8561 \\
	&ML 3$\sigma$& 0.0958 & 0.0188 & 0.0965  & 0.0201 & {\bf 0.0977}  & 0.0236 & {\bf 0.0949}  & 0.0330 & -0.0045  & 0.6403 \\
	$\phi_1=0.100$&BIP MM & {\bf 0.0980} & 0.0244 & 0.0977  & 0.0244 & 0.1036  & 0.0272 & 0.1135  & 0.0485& 0.0797  & 0.2488\\
	&BIP $\tau$ $c_{1,\mathrm{rob}}$ & 0.0956 & 0.0206 & 0.0967  & 0.0215 & 0.1045  & 0.0263 & 0.1135  & 0.0485& {\bf 0.0803}  & 0.2475\\
	&BIP $\tau$ $c_{1,\mathrm{eff}}$ & 0.0958 & 0.0188 & 0.0972  & 0.0215 & 0.1062  & 0.0271& 0.1280  & 0.0644& 0.0370  & 0.6239\\
	&BIP $\tau$ init & 0.0959 & 0.0187 & {\bf 0.0978}  & 0.0212 & 0.1038  & 0.0282 & 0.1196  &  0.0630 & 0.0260  & 0.6374\\
	\midrule
	&ML & {\bf 1.6539} & 0.0207 & 1.6339  & 0.1136 & 1.3544  & 0.5024 & 0.8911 & 0.6289& 0.7407 & 0.6524\\
	&ML 3$\sigma$& 1.6541 & 0.0210 & {\bf 1.6555}  & 0.0224 & {\bf 1.6549}  & 0.0252 & {\bf 1.6434}  & 0.0338 & 1.1933 & 0.5861 \\
	$\phi_2=1.6600$ &BIP MM & 1.6517 & 0.0284 & 1.6303  & 0.0250 & 1.6169  & 0.0349 & 1.6049  &0.0638& {\bf 1.5321}  & 0.1979\\
	&BIP $\tau$ $c_{1,\mathrm{rob}}$ & 1.6526 & 0.0250 & 1.6323  &0.0244 & 1.6168  & 0.0345 & 1.6048  & 0.0638& 1.5314  & 0.1978\\
	&BIP $\tau$ $c_{1,\mathrm{eff}}$ & 1.6537 & 0.0209 & 1.6377  & 0.0257 & 1.6242  & 0.0404& 1.6036 & 0.0724& 1.0778  & 0.5320\\
	&BIP $\tau$ init & 1.6540 & 0.0208 & 1.6346  & 0.0242 & 1.6066  & 0.0382 & 1.5645  &  0.0684 & 1.0858  & 0.5371\\
	\midrule
	&ML & 0.0879 & 0.0178 & 0.0729 & 0.1166 & -0.0571  & 0.3544 & -0.2001  & 0.6868& -0.2942  & 0.6054\\
	&ML 3$\sigma$& 0.0879 & 0.0178 & {\bf 0.0885}  & 0.0191 & 0.0904  & 0.0238 & {\bf 0.0884}  & 0.0332 & -0.1453  & 0.4816\\
	$\phi_3=0.0930$ &BIP MM & 0.0885 & 0.0251 & 0.0870  & 0.0214 & {\bf 0.0936}  & 0.0250 & 0.0808  &0.0316& 0.0272  & 0.1962\\
	&BIP $\tau$ $c_{1,\mathrm{rob}}$ & {\bf 0.0892} & 0.0199 & 0.0882  &0.0199 & 0.0918  & 0.0250 & 0.0808  & 0.0316& 0.0271  & 0.1967\\
	&BIP $\tau$ $c_{1,\mathrm{eff}}$ & 0.0877 & 0.0178 & 0.0876  & 0.0192 & 0.0879  & 0.0257& 0.0722 & 0.0387& -0.1186  & 0.4468\\
	&BIP $\tau$ init & 0.0879 & 0.0177 & 0.0862  & 0.0192 & 0.0885  & 0.0238 & 0.0874  &  0.0347 & {\bf 0.1071}  & 0.4402\\
	\midrule
	&ML & 0.8578 & 0.0197 & 0.8456  & 0.1124 & 0.6229  & 0.4511 & 0.2757  & 0.5881& 0.2800  & 0.5458\\
	&ML 3$\sigma$& 0.8580 & 0.0199 & {\bf 0.8590}  & 0.0224 & {\bf 0.8591}  & 0.0254 & {\bf 0.8580}  & 0.0359 & 0.5339  & 0.5154 \\
	$\phi_4=0.8649$ &BIP MM & 0.8572 & 0.0320 & 0.8415  & 0.0299 & 0.8215  & 0.0381 & 0.8082  &0.0792& 0.7606  & 0.1478\\
	&BIP $\tau$ $c_{1,\mathrm{rob}}$ & {\bf 0.8580} & 0.0245 & 0.8344  &0.0266 & 0.8171  & 0.0360 & 0.8082  & 0.0792& {\bf 0.7606}  & 0.1476\\
	&BIP $\tau$ $c_{1,\mathrm{eff}}$ & 0.8579 & 0.0203 & 0.8409  & 0.0267 & 0.8271  & 0.0428& 0.8033 & 0.0962& 0.4203  & 0.4032\\
	&BIP $\tau$ init & 0.8578 & 0.0199 & 0.8355  & 0.0243 & 0.8036  & 0.0396 & 0.7516  &  0.0868 & 0.4319  & 0.4094\\
	\midrule
	&ML & 0.0189 & 0.0427 & 0.0677  & 0.1519 & {\bf 0.0231}  & 0.4417 & {\bf 0.0585}  & 0.8319& -0.0594 & 0.8610\\
	&ML 3$\sigma$& {\bf 0.0199} & 0.0445 & 0.0391 & 0.0425 & 0.0540  & 0.0454 & 0.0768  & 0.0620 & -0.0202  & 0.6457\\
	$\theta_1=0.0226$ &BIP MM & 0.0188 & 0.0471 & 0.0382  & 0.0468 & 0.0581  & 0.0505 & 0.0874  & 0.0673& 0.0534  & 0.2451\\
	&BIP $\tau$ $c_{1,\mathrm{rob}}$ & 0.0191 & 0.0439 & 0.0387  & 0.0451 & 0.0581  & 0.0499 & 0.0874  & 0.0673& 0.0528  & 0.2461\\
	&BIP $\tau$ $c_{1,\mathrm{eff}}$ & 0.0187 & 0.0427 & {\bf 0.0371}  & 0.0433 & 0.0591  & 0.0474& 0.1001  & 0.0800& {\bf 0.0169}  & 0.6307\\
	&BIP $\tau$ init & 0.0187 & 0.0428 & 0.0382 & 0.0429 & 0.0620  & 0.0475 & 0.0966  &  0.0813 & 0.0111  & 0.6448\\
	\midrule
	&ML & 0.8156 & 0.0428 & 1.4043  & 0.1167 & 1.2001  & 0.5005 & {\bf 0.8150}  & 0.6334& 0.6882  & 0.6599\\
	&ML 3$\sigma$& 0.8260 & 0.0453 & 1.0068  & 0.0751 & 1.1269  & 0.0741 & 1.3654  & 0.0719 & 1.0539  & 0.5808\\
	$\theta_2=0.8175$ &BIP MM & {\bf 0.8171} & 0.0463 & 0.8504  & 0.0578 & 0.8780  & 0.0786 & 0.9739  & 0.1058& 1.1099  & 0.2089\\
	&BIP $\tau$ $c_{1,\mathrm{rob}}$ & 0.8151 & 0.0459 & {\bf 0.8520}  & 0.0568 & 0.8816  & 0.0786 & 0.9739  & 0.1058& 1.1105  & 0.2093\\
	&BIP $\tau$ $c_{1,\mathrm{eff}}$ & 0.8148 & 0.0425 & 0.8513  & 0.0521 & {\bf 0.8683}  & 0.0837& 1.0315  & 0.1281&  0.9044  & 0.5238\\
	&BIP $\tau$ init & 0.8151 & 0.0427 & 0.8562  & 0.0552 & 0.8831  & 0.0852 & 1.0706  &  0.1250 & {\bf 0.8974}  & 0.5340\\
	\midrule
	&ML & 0.0530 & 0.0461 & 0.0674  & 0.1067 & -0.0498  & 0.3111 & -0.1909 & 0.6486& -0.2856  & 0.5824\\
	&ML 3$\sigma$& 0.0534 & 0.0467 & 0.0639  & 0.0424 & 0.0738  & 0.0466 & 0.0906  & 0.0591 & -0.1388  & 0.4203 \\
	$\theta_3=0.0595$ &BIP MM & {\bf 0.0540} & 0.0479 & 0.0625  & 0.0499 & 0.0701  & 0.0552 & 0.0745  & 0.0646& {\bf 0.0395}  & 0.1199\\
	&BIP $\tau$ $c_{1,\mathrm{rob}}$ & 0.0538 & 0.0473 & 0.0387  & 0.0485 & 0.0713  & 0.0549 & 0.0745  & 0.0646& 0.0397  & 0.1191\\
	&BIP $\tau$ $c_{1,\mathrm{eff}}$ & 0.0529 & 0.0460 & {\bf 0.0606}  & 0.0458 & {\bf 0.0698}  & 0.0527& {\bf 0.0662}  & 0.0700& 0.1098  & 0.3731\\
	&BIP $\tau$ init & 0.0528 & 0.0461 & 0.0613  & 0.0479 & 0.0716  & 0.0553 & 0.0770  &  0.0770 & 0.0948  & 0.3656\\
	\midrule
	&ML & {\bf 0.0733} & 0.0371 & 0.6128 & 0.1146 & 0.5031 & 0.4012 & 0.2551 & 0.5656& {\bf 0.2676} & 0.5391\\
	&ML 3$\sigma$& 0.0819 & 0.0388 & 0.2349  & 0.0642 & 0.3424  & 0.0705 & 0.5661  & 0.0761 & 0.4389  & 0.4663 \\
	$\theta_4=0.0764$ &BIP MM & 0.0720 & 0.0373 & {\bf 0.0965}  & 0.0503 & 0.1211  & 0.0668 & {\bf 0.2192}  & 0.1012& 0.3856  & 0.1269\\
	&BIP $\tau$ $c_{1,\mathrm{rob}}$ & {\bf 0.0733} & 0.0364 & 0.0978  & 0.0494 & 0.1221  & 0.0654 & {\bf 0.2192}  & 0.1012& 0.3857  & 0.1268\\
	&BIP $\tau$ $c_{1,\mathrm{eff}}$ & 0.0729 & 0.0372 & 0.1005  & 0.0514 & 0.1163  & 0.0723& 0.2694  & 0.1254& 0.3194 & 0.3555\\
	&BIP $\tau$ init & 0.0726 & 0.0372 & 0.0997  & 0.0541 & {\bf 0.1146}  & 0.0758 & 0.2783  &  0.1321 & 0.3159  & 0.3798\\
\bottomrule
	\end{tabular}
\label{table:example_arma44}
\end{table*}

\bibliographystyle{ieee}

\end{document}